\documentclass[twocolumn]{aastex631}

\let\tablenum\relax
\usepackage{siunitx}

\defcitealias{Mendoza_2022}{M22}
\defcitealias{Howard_2022}{H22}

\received{2024 January 12}
\revised{2024 February 23}
\accepted{2024 February 23}
\submitjournal{\apjs}

\shorttitle{Flares in compact stars from TESS}
\shortauthors{Xing et al.}

\begin{document}

\title{Flares hunting in hot subdwarf and white dwarf stars from Cycles 1-5 of TESS photometry}

\author[0009-0003-8858-2833]{Keyu Xing}
\affiliation{Institute for Frontiers in Astronomy and Astrophysics, Beijing Normal University, Beijing~102206, P.~R.~China}
\affiliation{Department of Astronomy, Beijing Normal University, Beijing~100875, P.~R.~China}
\affiliation{International Centre of Supernovae, Yunnan Key Laboratory, Kunming, 650216, P.~R.~China}

\author[0000-0002-7660-9803]{Weikai Zong}
\affiliation{Institute for Frontiers in Astronomy and Astrophysics, Beijing Normal University, Beijing~102206, P.~R.~China}
\affiliation{Department of Astronomy, Beijing Normal University, Beijing~100875, P.~R.~China}
\affiliation{International Centre of Supernovae, Yunnan Key Laboratory, Kunming, 650216, P.~R.~China}

\author[0000-0002-1295-8174]{Roberto Silvotti}
\affiliation{INAF-Osservatorio Astrofisico di Torino, strada dell'Osservatorio 20, I-10025 Pino Torinese, Italy}

\author[0000-0001-8241-1740]{Jian-Ning Fu}
\affiliation{Institute for Frontiers in Astronomy and Astrophysics, Beijing Normal University, Beijing~102206, P.~R.~China}
\affiliation{Department of Astronomy, Beijing Normal University, Beijing~100875, P.~R.~China}

\author[0000-0002-6018-6180]{Stéphane Charpinet}
\affiliation{Institut de Recherche en Astrophysique et Planétologie, CNRS, Université de Toulouse, CNES, 14 Avenue Edouard Belin, F-31400 Toulouse, France}

\author[0000-0003-3816-7335]{Tianqi Cang}
\affiliation{Institute for Frontiers in Astronomy and Astrophysics, Beijing Normal University, Beijing~102206, P.~R.~China}
\affiliation{Department of Astronomy, Beijing Normal University, Beijing~100875, P.~R.~China}
\affiliation{International Centre of Supernovae, Yunnan Key Laboratory, Kunming, 650216, P.~R.~China}

\author[0000-0001-5941-2286]{J. J. Hermes}
\affiliation{Department of Astronomy \& Institute for Astrophysical Research, Boston University, 725 Commonwealth Avenue, Boston, MA 02215, USA}

\author[0000-0002-7468-3612]{Xiao-Yu Ma}
\affiliation{Institute for Frontiers in Astronomy and Astrophysics, Beijing Normal University, Beijing~102206, P.~R.~China}
\affiliation{Department of Astronomy, Beijing Normal University, Beijing~100875, P.~R.~China}

\author[0009-0004-1774-7167]{Haotian Wang}
\affiliation{Institute for Frontiers in Astronomy and Astrophysics, Beijing Normal University, Beijing~102206, P.~R.~China}
\affiliation{Department of Astronomy, Beijing Normal University, Beijing~100875, P.~R.~China}

\author[0000-0002-6457-4607]{Xuan Wang}
\affiliation{Institute for Frontiers in Astronomy and Astrophysics, Beijing Normal University, Beijing~102206, P.~R.~China}
\affiliation{Department of Astronomy, Beijing Normal University, Beijing~100875, P.~R.~China}

\author[0000-0001-6832-4325]{Tao Wu}
\affiliation{Yunnan Observatories, Chinese Academy of Sciences, 396 Yangfangwang, Guandu District, Kunming, 650216, P.~R.~China}
\affiliation{Key Laboratory for the Structure and Evolution of Celestial Objects, Chinese Academy of Sciences, 396 Yangfangwang, Guandu District, Kunming, 650216, P.~R.~China}
\affiliation{Center for Astronomical Mega-Science, Chinese Academy of Sciences, 20A Datun Road, Chaoyang District, Beijing, 100012, P.~R.~China}
\affiliation{University of Chinese Academy of Sciences, Beijing 100049, P.~R.~China}
\affiliation{International Centre of Supernovae, Yunnan Key Laboratory, Kunming, 650216, P.~R.~China}

\author[0000-0002-6868-6809]{Jiaxin Wang}
\affiliation{College of Science, Chongqing University of Posts and Telecommunications, Chongqing 400065, P.~R.~China}

\correspondingauthor{Weikai Zong}
\email{weikai.zong@bnu.edu.cn}

\begin{abstract}

Stellar flares are critical phenomena on stellar surfaces, which are closely tied to stellar magnetism. While extensively studied in main-sequence (MS) stars, their occurrence in evolved compact stars, specifically hot subdwarfs and white dwarfs (WDs), remains scarcely explored. Based on Cycles 1-5 of TESS photometry, we conducted a pioneering survey of flare events in \num{\sim12000} compact stars, corresponding to \num{\sim38000} light curves with 2-minute cadence. Through dedicated techniques for detrending light curves, identifying preliminary flare candidates, and validating them via machine learning, we established a catalog of 1016 flares from 193 compact stars, including 182 from 58 sdB/sdO stars and 834 from 135 WDs, respectively. However, all flaring compact stars showed signs of contamination from nearby objects or companion stars, preventing sole attribution of the detected flares. For WDs, it is highly probable that the flares originated from their cool MS companions. In contrast, the higher luminosities of sdB/sdO stars diminish companion contributions, suggesting that detected flares originated from sdB/sdO stars themselves or through close magnetic interactions with companions. Focusing on a refined sample of 23 flares from 13 sdB/sdO stars, we found their flare frequency distributions were slightly divergent from those of cool MS stars; instead, they resemble those of hot B/A-type MS stars having radiative envelopes. This similarity implies the flares on sdB/sdO stars, if these flares did originate from them, may share underlying mechanisms with hot MS stars, which warrants further investigation.

\end{abstract}


\section{Introduction} \label{sec:intro}

Stellar flares are abrupt and intense phenomena that manifest as a rapid increase in luminosity across a broad wavelength coverage \citep{Benz_2010}, posing significant impacts on the habitability of orbiting exoplanets \citep[e.g.,][]{Vida_2017}. The underlying mechanisms that generate stellar flares are believed to be analogous to those of solar flares \citep{Davenport_2016}, closely related to magnetic activity on stellar surfaces. They are triggered by the sudden release of magnetic energy through the reconnection of twisted magnetic field lines within the stellar atmosphere. In general, stars with deeper convective layers are capable of generating more vigorous dynamo mechanisms \citep{Charbonneau_2010}, leading to stronger surface magnetic fields and higher frequencies of flares. This relationship manifests as an observed trend where the incidence of flare stars increases progressively from F-type to M-type main-sequence stars \citep{Yang_2019}.
 
Incapable of resolving details of stellar surfaces, studies of stellar flares rely on time-resolved photometric or spectroscopic observations \citep{Walkowicz_2011}. Before the era of space-borne photometry, various types of observations of flares, for instance, radio waves \citep{Bastian_1988}, X-rays \citep{Schmitt_1993}, and optical spectroscopy \citep{Pettersen_1989}, offer different insights to their properties but lacking in continuous data coverage for their temporal evolution. While time-resolved spectroscopy enables detailed plasma diagnostics and detection of mass ejections of single flare events \cite[e.g.][]{Namekata_2021}, continuous photometric monitoring provides a broader view, allowing for statistical analyses of flare frequencies and energies over extended periods. The Kepler, K2, and TESS missions \citep{Koch_2010, Howell_2014, Ricker_2015}, collecting high-quality and long consecutive photometry, have fertilized the field of flare studies, exposing various properties and correlations of flares to their hosting stellar types across the HR diagram \citep[e.g.,][]{Davenport_2016, Gunther_2020, Yang_2023}.

Despite these advances, the knowledge of evolved compact stars, specifically hot subdwarfs and white dwarfs, remains very limited. Hot subdwarfs are evolved stars with B to O spectral types (sdB/sdO stars), burning helium in the core with a thin hydrogen envelope. They are located at the blue end of the horizontal branch, thus also known as extreme horizontal branch (EHB) stars \citep{Heber_2009}. White dwarfs (WDs) are remnants of $\sim97\%$ stars in the Milky Way when they cease nuclear fusion, resulting in a degenerate core typically composed of carbon and oxygen and cooling down for the rest of their lives \citep{Saumon_2022}. Contrary to cool MS stars, hot compact stars do not present deep convective envelopes, which prevents the generation of surface magnetic fields through dynamo processes.

Nevertheless, strong magnetism has been claimed in WDs and sdB/sdO stars \citep{Bagnulo_2021, Vos_2021, Pelisoli_2022}. Now that WDs are found with magnetism whose strength is in a large range from a few kG up to 1000 MG \citep[e.g.,][]{Landstreet_2019}. Their magnetism is explained by fossil mechanisms or the conservation of magnetic flux. Recent observations have shown the presence of spots on an increasing number of WDs \citep[e.g.,][]{Kilic_2015, Hoard_2018, Hermes_2021}, implying that flares may be able to occur on these stars, as both phenomena are closely tied to magnetic fields. Moreover, spot modulation is empirically not restricted to purely convective or strongly magnetic WDs \citep{Hermes_2017}, thereby opening the possibility for flare occurrences on WDs having a radiative envelope or relatively weak magnetic field. On the other hand, magnetism in sdB/sdO stars is very rare \citep{Landstreet_2012} and only a few of them were claimed to have detectable magnetic field \citep{Vos_2021, Pelisoli_2022}. Therefore, whether these magnetic compact stars can present flare events is an open interesting question to understand the mechanism between flare and magnetism.

However, it is a challenge to search for flare events in sdB/sdO stars and WDs since their sample is relatively small due to their low brightness and flare hunting needs intensive photometric monitoring. TESS mission provides the first opportunity for such research as it collected extensive photometry for a large fraction of sdB/sdO stars and WDs brighter than $G<16$ \citep{Ricker_2015, Charpinet_2019}. TESS enables the investigation of various types of brightness variation in those compact stars, for instance, brightness modulation from different binary effects \citep[see, e.g.,][]{Schaffenroth_2022} and rapid variation from gravity and pressure pulsations \citep[see, e.g.,][]{Romero_2022, Baran_2023}. Here we initiate the pioneering survey with an aim that tries to find flare events in sdB/sdO stars and WDs.

In line with this study, the light curves of compact stars can present complex variability, which brings difficulty in flare identification. For example, \citet{Pietras_2022} excluded stars with spectral types earlier than F1 when detecting stellar flares in TESS photometry, due to their diverse and rapid brightness variations that can easily cause false positives. Moreover, previous literature has reported that sporadic outbursts in cool hydrogen-atmosphere pulsating white dwarfs (DAVs) \citep{Bell_2015, Bell_2016, Hermes_2015}, which can be explained by limit cycles due to sufficiently resonant 3-mode couplings \citep{Luan_2018}. \citet{Scaringi_2022} also reported localized thermonuclear bursts in three accreting white dwarfs. Besides, self-lensing flares of WDs with compact companions can produce periodic brightening events if viewed close to edge on \citep{Nir_2023}. These various outbursts have different characteristic profiles and durations compared to stellar flares. Careful inspection is therefore necessary to conclusively identify stellar flares amidst the complex variability in compact star light curves.

Although there are several existing methods to detect flare events automatically \citep[e.g.][]{Davenport_2016, Doorsselaere_2017, Gunther_2020, Ilin_2021}, they generally rely on simple outlier detection in flattened light curves without considering event morphology. As a result, they struggle to distinguish flares from artifacts or other rapid brightening events. \citet{Yang_2019} found serious contamination of previous flare catalogs by various pulsators, rapid rotators, and transits. Other methods for detecting flares rely on machine learning algorithms \citep[e.g.,][]{Feinstein_2020}, but these may perform poorly when analyzing light curves with fast and complex brightness modulations. Therefore, detrending the light curves of compact variable stars correctly and distinguishing intrinsic flares from other transient events requires dedicated techniques for flare identification and validation.

The paper is structured as follows. In Section \ref{sec:photometry}, we describe the TESS photometry collected on our concerning compact stars. Section \ref{sec:detection} provides details of our methods for preliminary identification of flare candidates, including a new approach designed to address short-term periodic variations that enhances our ability to identify flares in compact stars. We also describe several indispensable steps to exclude contamination from cataclysmic variables and solar system objects. Section \ref{sec:validation} then explains our process to validate the preliminary candidates using a Random Forest classifier trained on a series of simulated data. We present the resulting flare catalog and analyze the properties of these events in Section \ref{sec:results}, also considering potential contamination and establishing a refined sample of compact stars. Finally, Section \ref{sec:summary} provides a summary of our results and a discussion of their implications regarding flare production mechanisms in compact stars.

\section{TESS Photometry} \label{sec:photometry}

The Transiting Exoplanet Survey Satellite \citep[TESS;][]{Ricker_2015} was launched on 18 April 2018, and began observing in July 2018. After completing the primary mission in July 2020 and the first extended mission (EM1) in September 2022, TESS is currently conducting its second extended mission (EM2), which will last approximately three years. 

Both the primary mission and EM1 of TESS consist of two observation cycles each, while EM2 comprises three cycles (Cycles 1-2, 3-4, and 5-7, respectively). Cycles 1-3 each included 13 sectors, with Cycle~1 and 3 pointing on the southern ecliptic hemisphere and Cycle~2 on the northern hemisphere. Cycle~4, totaling 16 sectors, continued observations in the northern hemisphere and added several sectors along the ecliptic plane. The recently completed Cycle~5, consisting of 14 sectors, covered both the northern and southern ecliptic hemispheres. Each sector covers a \qtyproduct{24 x 96}{\degree} strip of the sky with an angular resolution of \qty{21}{arcsec.pixel^{-1}} over an observational period of \qty{\sim27}{\day}. Due to the overlaps among the sectors in each cycle, a portion of TESS targets were observed multiple times during the survey. Targets in the vicinity of the ecliptic poles were observed continuously for almost one year.

TESS is equipped with four \qty{10.5}{\cm} optical telescopes and four identical cameras having a red bandpass covering the wavelength range from \qtyrange{600}{1000}{\nm}. Each camera has a field of view of \qtyproduct{24 x 24}{\degree} and consists of four \qtyproduct{2 x 2}{\unit{k}} back-illuminated CCDs, which continuously read out at 2-second intervals for spacecraft guiding. These 2-second data are then stacked, compressed, and stored on the spacecraft until TESS reaches its perigee after every 13.7-day orbital period, at which point TESS downlinks the collected data to Earth.

During Cycles 1-5 (Sector 1-69), TESS observed \num{11618} compact stars, including 7505 WDs and 4113 sdB/sdO stars, proposed by the TESS Asteroseismic Science Consortium (TASC)\footnote{\url{https://tasoc.dk}} Working Group 8 (WG8), which focuses on variability in evolved compact stars \citep[see, e.g.,][]{Charpinet_2019, Bognar_2020}. There are \num{38102} light curves with 2-minute cadence that have been collected for these compact stars. We retrieve their light curves from the Mikulski Archive for Space Telescopes (MAST)\footnote{\url{https://archive.stsci.edu/tess/bulk\_downloads.html}}, which were processed through the standard pipeline of the Science Processing Operations Center (SPOC) \citep{Jenkins_2016}. Table \ref{tab:sectors} lists the detailed observations of these compact stars: nearly two-thirds of the targets were observed in one or two sectors, while about one-tenth were visited by more than five sectors. We analyze all these \num{38102} light curves and characterize flare events using the Pre-search Data Conditioning Simple Aperture Photometry (PDCSAP) flux, where common instrumental systematics has been removed using the co-trending basis vectors. In each light curve, all epochs with quality bit flags equal to 1, 2, 3, 4, 5, 6, 8, 10, 13, and 15 are masked out, which are the recommended quality flags in the TESS Data Product documentation\footnote{\url{https://outerspace.stsci.edu/display/TESS/2.0+-+Data+Product+Overview}}.

\begin{deluxetable}{Ccc}
\setlength{\tabcolsep}{12pt}
\tablecaption{Number of evolved compact stars and light curves observed in TESS Cycles 1-5 with data available from $N$ sectors ($N$ from 1 to $\geqslant 6$). \label{tab:sectors}}
\tablehead{
\colhead{Number of sectors} & \colhead{Targets} & \colhead{Light curves}
}
\startdata
1 & 3521 & 3521 \\
2 & 3401 & 6802 \\
3 & 2149 & 6447 \\
4 & 921 & 3684 \\
5 & 436 & 2180 \\
\geqslant 6 & 1190 & \num{15468} \\
\hline
{\rm Total} & \num{11618} & \num{38102} \\
\enddata
\end{deluxetable}

\section{Flare Candidates Detection} \label{sec:detection}

Flares are shown as consecutive positive outliers in light curves. To detect these outliers, we first detrend the light curves to remove astrophysical variability. Then, we identify groups of outliers in the detrended light curves that meet certain criteria, which are regarded as preliminary flare candidates since their profiles have not yet been considered. We exclude the light curves of known cataclysmic variables (CVs) due to their complexity and irregularity. We also inspect whether each candidate is caused by a solar system object (SSO) encounter event.

\subsection{Detrending Algorithm and Window Length}

For many studies on flare detection, the Savitzky-Golay filter \citep{Savitzky_1964} has been adopted to detrend the light curve by various groups \citep[e.g.][]{Ilin_2021}. However, this filter is cadence-based and cannot function as designed on time series with observational gaps. \citet{Hippke_2019} compared the relative performance of various detrending algorithms for transit discovery. They concluded that generally it is optimal to use a time-windowed slider with an iterative robust location estimator based on Tukey’s biweight \citep{Mosteller_1977}. This result can also be applied to flare detection since transits and flares are both signals constituting small segments of the light curves. Meanwhile, the Savitzky-Golay filter easily overfits the transit signals, which reduces their depth and warps up the detrended light curve around the transit. This will introduce extra outliers and bring some false alarm detection of flares \citep[see Figure 1 \& 11 in][]{Hippke_2019}. Therefore, instead of using the Savitzky-Golay filter, we adopt the biweight filter implemented in \texttt{W\={o}tan} \citep{Hippke_2019}, an open-source Python package for time-series smoothing, to detrend the light curves. 

\begin{figure*}
\gridline{\fig{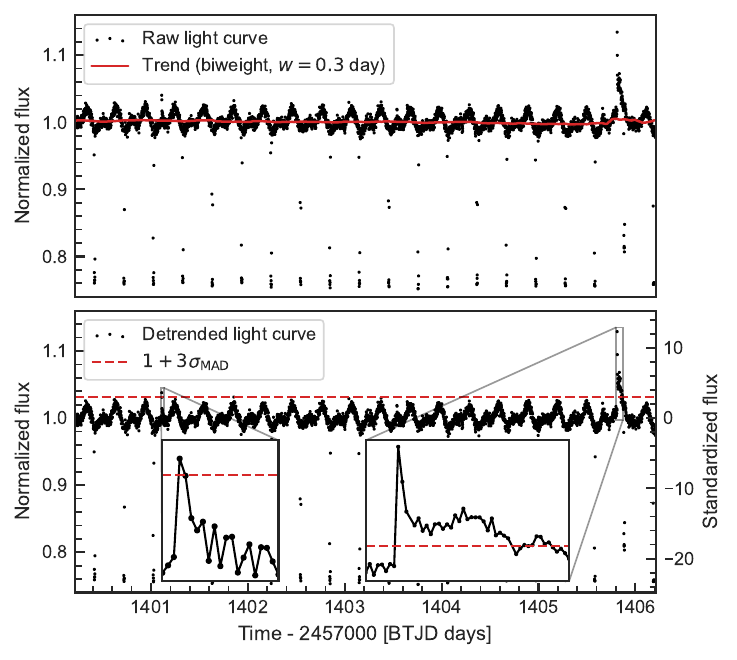}{\columnwidth}{(a)} \hspace{-1cm}
          \fig{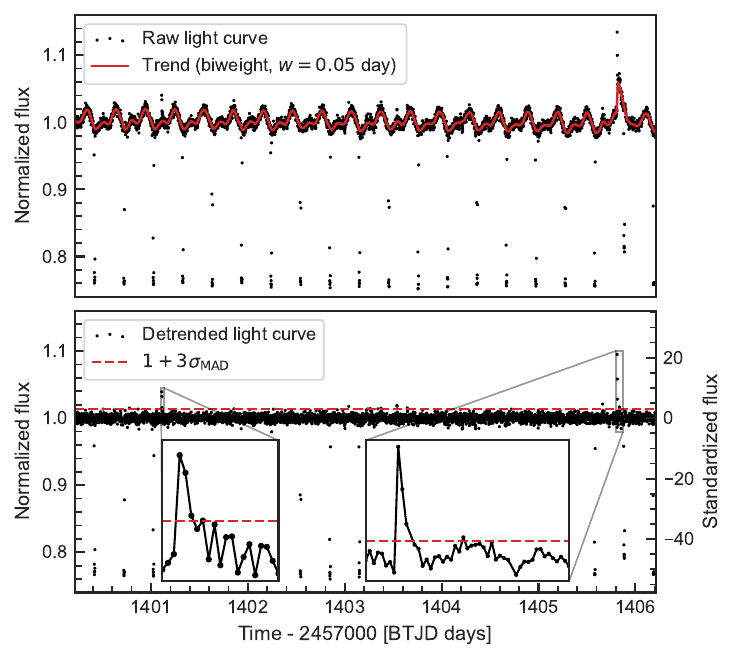}{\columnwidth}{(b)}}
\caption{Detrending the light curve of TIC 219244444 \citep[RR Cae, a WD eclipsed by an M-dwarf every \qty{\sim0.3}{\day};][]{Maxted_2007} in TESS Sector 3 using the biweight filter with two different window lengths: (a) long of \qty{0.3}{\day}, (b) short of \qty{0.05}{\day}. Top: raw flux (black points) and trend flux (red solid line). Bottom: detrended flux (black points), $3\sigma_{\rm MAD}$ from the median (red dashed line) and zoom windows of one small flare (left) and one large flare (right). The large flare is overfitted with the short window, while the small flare cannot be detected with the long window. Only a 6-day segment of the light curve is shown in this figure.}
\label{fig:window}
\end{figure*}

Next, we need to correctly determine the window length for the biweight filter before detrending. A narrow window can remove the stellar variability effectively but may overfit the large flares, which introduces underestimations of their duration and amplitude, as shown in Figure \ref{fig:window}(b). To avoid overfitting flares, the window length should be several times longer than their durations. A recent work by \citet[][hereafter H22]{Howard_2022} identified 3792 flares from 226 M-dwarfs based on 20-s cadence photometry from TESS Cycle~3, concluding that only \qty{\sim5}{\percent} of the detected flares have a duration longer than \qty{0.1}{\day}. We thus set the window length to \qty{0.3}{\day} in the application of the biweight filter, which is considerably long enough for the durations of most flares. Such window length can also effectively filter out other transient events with typical timescales over \qty{0.3}{\day} in compact stars, such as pulsation outbursts in DAVs \citep{Bell_2017}.

\subsection{Detrending Procedure}
Flares typically last from minutes to hours, rarely longer than \qty{1}{\day} \citep{Ilin_2022}. In some cases, the light curves of compact stars exhibit periodic variations, on a timescale order $\lesssim$\qty{1}{\day}, induced by factors such as pulsations or binary effects. Such variations cannot be easily removed if a wide window is chosen to preserve large flares during the detrending process, as shown in Figure \ref{fig:window}(a). In addition, it can result in the failure to detect flares with amplitudes comparable to those periodic variations. We thus have to implement an automated procedure for detrending light curves, with a particular focus on identifying and removing these short-term periodic variations.

We firstly normalize the light curves by dividing them by the total median and compute the Lomb-Scargle periodogram \citep{Lomb_1976, Scargle_1982}. If the period of the highest peak, $P$, is in the range of \qtyrange{0.05}{2}{\day} and with significant confidence (we arbitrarily adopt 4$\times$ the median noise level), then this light curve will be considered to have short-period variability, otherwise it will be directly detrended. In some special cases, instead of the period of the maximum power signal, its harmonics represent the true period of brightness variation. Therefore, the light curve will be folded with $P$, $2P$, and $4P$, respectively. We then apply a median filter with a sliding window of $P_{\rm fold}/50$ to each folded light curve to generate models of the short-period variability, where $P_{\rm fold}$ is the folded period of the light curve (Figure \ref{fig:model}). After evaluating the standard deviations of the residuals of all models, the optimum model will be selected and subtracted from the light curve. Lastly, the differential light curves will be detrended using the biweight filter with a window length of \qty{0.3}{\day}. Both large and small flare profiles are preserved and can be easily detected through this procedure (Figure \ref{fig:detrend}).

\subsection{Removing Detached Eclipses} \label{sec:removing_eclipses}

\begin{figure*}
\epsscale{1.2}
\plotone{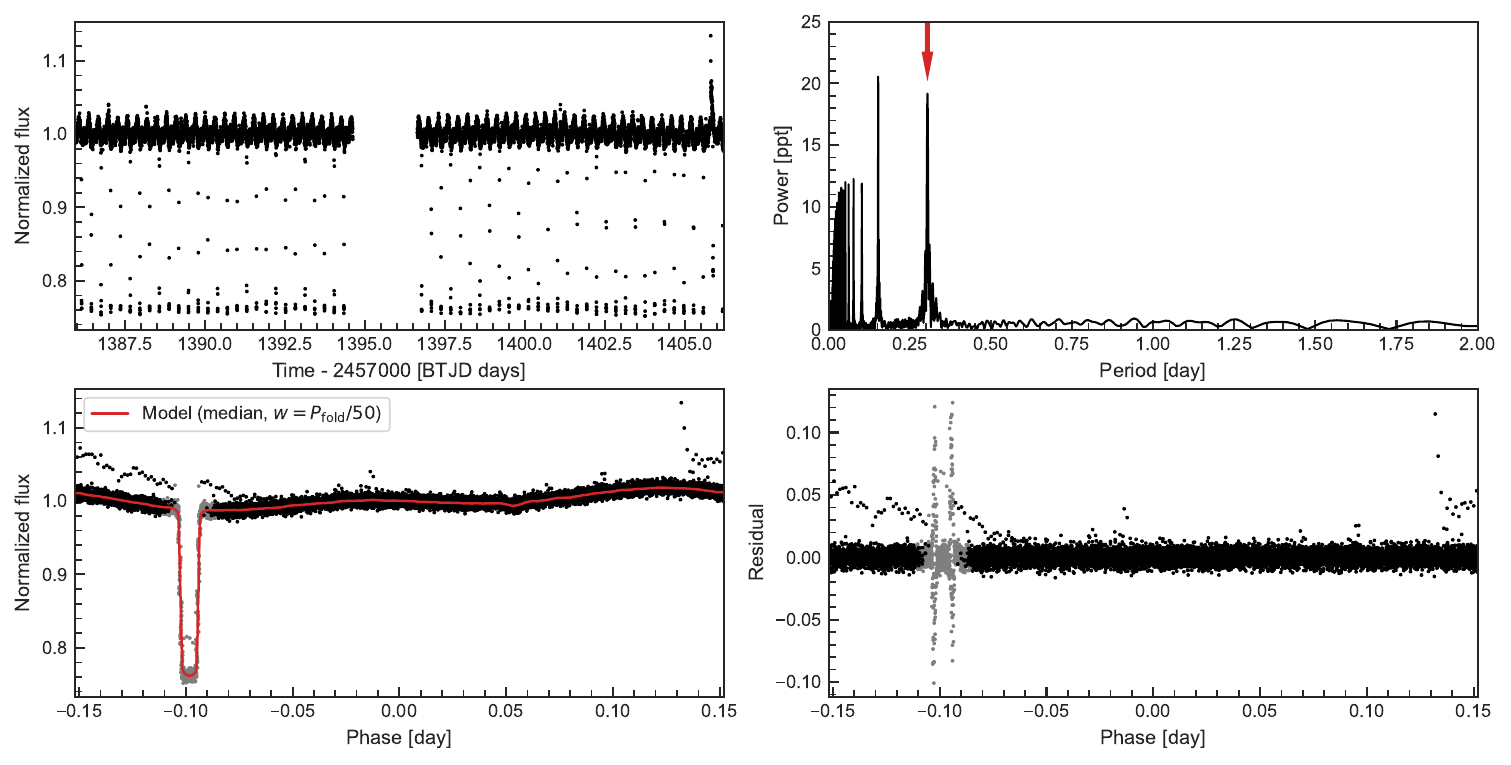}
\caption{An example of generating the model of the short-period variability in the light curve of TIC 219244444. Top: The light curve of TIC 219244444 in TESS Sector 3 (left) and its Lomb–Scargle periodogram (right). The red arrow indicates the period used to fold the light curve ($P_{\rm fold}=0.304\,\rm{d}$), which is twice the period at the max power. Bottom: The folded light curve and its model (red solid line) obtained by applying the median filter (left) and the residual of the model (right). The epochs masked by the processing in Section \ref{sec:removing_eclipses} are marked with grey points. The model fits the short-period variability well except the ingress and egress of the eclipse.}
\label{fig:model}
\end{figure*}

\begin{figure}
\epsscale{1.2}
\plotone{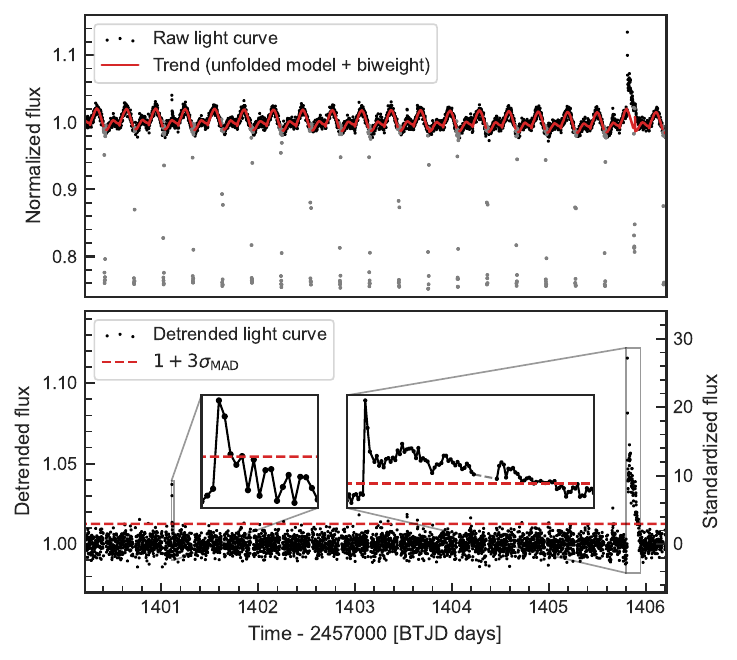}
\caption{The same as Figure \ref{fig:window}, but using our detrending procedure instead of directly using the biweight filter. The trend flux (red solid line) is the combination of the unfolded model of the short-period variability and the trend flux obtained using the biweight filter afterwards. The grey points in the top panel are removed (see Section \ref{sec:removing_eclipses}) in the detrended light curve, which causes the missing data (grey dashed line) in the right zoom window of the bottom panel.}
\label{fig:detrend}
\end{figure}

The aforementioned procedure can effectively detrend most light curves, but it fails, in particular, for a few with detached eclipses. The folded models cannot correctly recover the periodic variations because detached eclipsing signals require many harmonics to be represented in Fourier transformation. To avoid this case, we therefore remove the eclipses before applying the biweight filter. The eclipses are masked in the light curve where more than three consecutive epochs are below $-3 \sigma_{\rm MAD}$ from the median before subtracting the folded model. Here $\sigma_{\rm MAD}$ is a robust standard deviation calculated using the median absolute deviation (MAD), given by
\begin{equation}
\rm \sigma_{MAD} \approx 1.4826 \, MAD,
\end{equation}
assuming that the data is normally distributed \citep{Huber_1981}. We then extend each masked segment on both sides until an epoch is above the median of the light curve, ensuring the entire eclipse is covered (Figure \ref{fig:model}).

\subsection{Searching for Flare Candidates} \label{sec:searching_candidates}

After detrending the light curve, we slide a 2-day window along it and calculate its rolling $\sigma_{\rm MAD}$ to estimate the local noise level. We then standardize the light curve by
\begin{equation}
	F_{\rm s} = \frac{F_{\rm d} - 1}{\sigma_{\rm MR}},
\end{equation}
where $F_{\rm d}$ is the normalized detrended light curve, $\sigma_{\rm MR}$ is the rolling $\sigma_{\rm MAD}$ and $F_{\rm s}$ is the standardized light curve. Standardization is a technique that transforms data to have a mean of 0 and a standard deviation of 1, commonly used when datasets have varying scales. In this case, we use standardized light curves, which represent the deviation of the flux measurements from the median in units of $\sigma_{\rm MR}$, to simplify identifying and validating preliminary flare candidates in subsequent analyses. 

To identify groups of outliers as preliminary flare candidates, we search for at least $N_1$ consecutive measurements above $N_2\times\sigma$ from the median, i.e., greater than $N_2$ in the standardized light curve. Lower values of $N_{1,2}$ recover more low-amplitude flares but increase the number of false positives due to artifacts, and vice versa. We empirically select a threshold of $N_1=2$ and $N_2=3$, compromising between the efficiency of recovering flare events and minimizing false positives.

Although the above criteria are sufficient for identifying flare candidates, the start of their rising phases and the end of their decaying phases may be cut off if they lie below the $3\sigma$ threshold. We therefore extend the profiles of flare candidates, similar to the extension process in Section \ref{sec:removing_eclipses}, to provide a more accurate estimation of their parameters. Considering the sharp rise and gradual decay of flare events, we extend the left and right sides of each candidate event until one and two consecutive epochs, respectively, fall below $\sigma_{\rm MAD}$ from the median, i.e., less than 1 in the standardized light curve.

\subsection{Rejection of CVs and SSO Encounters}

Cataclysmic variables (CVs) have also been proposed in the TASC~WG8 target list as they consist of a WD primary and a mass transferring secondary. However, it is difficult to identify flare events in the light curves of CVs since they exhibit rich variable properties across different time scales from seconds to millennia \citep{Bruch_2022}. The majority of outbursts are not triggered by flare events (e.g. superhumps), leading to severe pollution in our identification. We thus have to disregard those light curves by querying the type of each compact star from the SIMBAD astronomical database\footnote{\url{https://simbad.cds.unistra.fr/simbad/}} \citep{Wenger_2000}.

\begin{figure*}
\epsscale{1.2}
\plotone{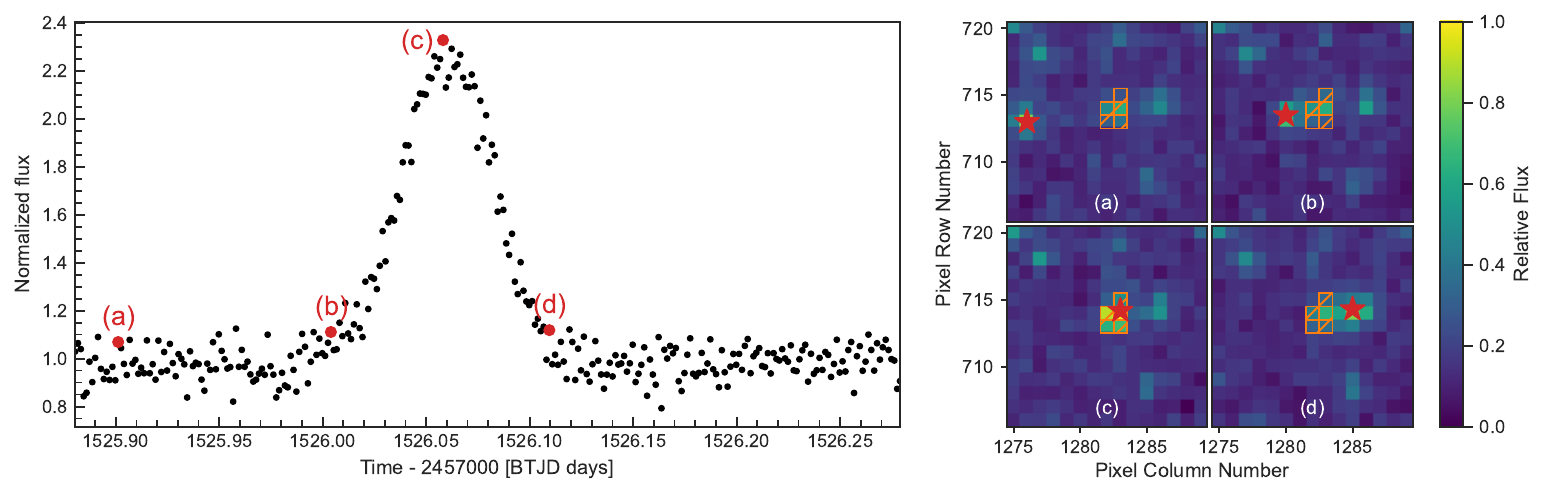}
\caption{An SSO encounter event found in the light curve of TIC 275308213 in TESS Sector 8. The light curve shown in the left panel is the photometry result using the aperture shown as orange region in the right panel, where presents four frames of the target pixel file during the process that an SSO, marked by red star symbol, encounters the target star. The corresponding fluxes of these moments are also marked in the light curve.}
\label{fig:encounter}
\end{figure*}

When a small, foreground solar system object (SSO) (e.g. asteroid, comet) moves across the aperture mask of a target, it tends to cause a symmetrical profile of increasing flux on the light curve as shown in Figure \ref{fig:encounter}. This phenomenon has been extensively discussed by \citet{Pal_2018}. The SSO encounter events can be easily misidentified as flare candidates and hence need to be excluded. We use the SkyBoT\footnote{\url{https://vo.imcce.fr/webservices/skybot/}} service \citep{Berthier_2006} to identify the false signals caused by SSO encounters, which has been implemented in the Kepler/K2 images \citep{Berthier_2016}. For each flare candidate, we perform a cone search in SkyBoT with a radius of 8 TESS pixels, corresponding to \qty{2.8}{\arcmin}, at the peak time, which returns a list of known SSOs located in the vicinity of the target. However, a very faint SSO will have little impact on the light curve even if it is identified with flare events \citep[see e.g.][]{Gunther_2020}. We thus also query the visual magnitude of the SSOs from the JPL Horizons system\footnote{\url{https://ssd.jpl.nasa.gov/horizons/}} \citep{Giorgini_2001} and only remove those with $V_{\rm mag}<19$, which will retain the real flares occurring on the targets passing by faint SSOs.

\section{Flare Candidates Validation} \label{sec:validation}

While the criteria described above are effective in identifying preliminary flare candidates, they may also capture other types of transient events or artifacts, resulting in a considerable number of false positives. Despite our efforts to minimize them through the aforementioned steps, some non-flare events may still exist due to various reasons such as poor data quality, uncategorized CVs in SIMBAD, and targets affected by unknown SSOs. As a result, a more rigorous validation process is required to eliminate non-flare events from the preliminary flare candidates.

Owing to the substantial number of preliminary candidates, we adopt Random Forest \citep[RF;][]{Breiman_2001}, a supervised machine learning algorithm, to automatically evaluate the confidence level of each candidate. We generate batches of flare and non-flare events through a series of simulations, which serve as the training inputs for the RF classifier. The classifier then evaluates each candidate and returns a confidence probability for it being an intrinsic flare event. Candidates with a probability that exceeds a certain confidential threshold are accepted as validated candidates.

\subsection{Random Forest}

Random Forest (RF) is an ensemble machine learning method that can be used for both regression and classification problems. It is constructed by combining multiple decision trees, with each tree being trained on a different subsample of the training set generated by bootstrap aggregation, also known as bagging \citep{Breiman_1996}. For classification tasks, the RF classifier outputs the class probabilities of an input sample by averaging the predicted class probabilities from each tree in the forest. Unlike a single decision tree, which can be unstable and sensitive to noise, RF combines the results of numerous weakly correlated trees, providing more accurate and robust predictions. 

Due to its high accuracy, low variance, and ease of application, RF has been widely used for various classification tasks \citep[e.g.][]{Brink_2013, Wyrzykowski_2015, Wyrzykowski_2016, Godines_2019}. Previous literature has also reported that RF outperforms other machine learning methods when classifying variable stars \citep[see, e.g.,][]{Richards_2011, Brink_2013, Pashchenko_2018}. Given these successful applications and outstanding performance in comparisons, we choose to apply RF to validate the preliminary flare candidates. For this research, we use \texttt{scikit-learn} \citep{scikit-learn}, an open source Python package for machine learning, to build the RF classifier.

\subsection{Training Set}

As a supervised learning method, RF requires a training set that is composed of objects with known labels. We choose to construct the training set from simulations rather than relying on data from previous surveys. Using simulations enables the creation of a comprehensive and accurate training set with minimal label contamination because it allows for precise control of input parameters and the generation of a broader variety of flare events, which might be underrepresented in previous survey data due to incomplete samples of low-energy flares \citep{Feinstein_2020}.

\subsubsection{Flare Model}

\citet{Davenport_2014} generated an empirical flare template using classical (single peak) flares discovered in all 11 months of 1-minute cadence data for GJ 1243, an active M4 star, available from \textit{Kepler} Data Release 23. This flare template has been widely used for constructing light curves of flare stars \citep[e.g.][]{Gunther_2020}. Recently, \citet[][hereafter M22]{Mendoza_2022} reanalyzed the same data for GJ 1243 using the \textit{Kepler} Data Release 25, where the light curve processing was improved. They generated an updated analytic and continuous flare template, addressing the limitations of the flare template in \citet{Davenport_2014}. We thus adopt the flare template in \citetalias{Mendoza_2022} to generate the simulated flare events here.

The \citetalias{Mendoza_2022} flare template used the convolution of a Gaussian and a double exponential to model the morphology of the flares, which can be parameterized with three variables: amplitude, the full time width at half the maximum of the flux \citep[FWHM, also known as $t_{1/2}$ in][]{Kowalski_2013} and center time \citep[similar to $t_{\rm peak}$ in][the moment the flare peaks]{Davenport_2014}. The flare template was normalized to a relative flux scale, ranging from 0 (before and after the flare occurs) to $A$, the amplitude (at the flare peak). By changing the parameters from specific distributions, we can generate various simulated flare events using this flare template.

\subsubsection{Parameters Fitting for Simulated Flares}

To create simulated flares that accurately represent real ones as close as possible, we use observed data to fit the distributions of flare parameters. By drawing parameters from these fitted distributions, we can generate simulated flare events that are statistically representative of real ones. Since the center time does not affect profiles of flare, we completely concentrate on determining the distributions of its amplitude and FWHM.

We specifically choose to use the parameters of the \num{3792} M-dwarf flares listed in \citetalias{Howard_2022}, as they provide the photometric signal-to-noise ratio (S/N), $\sigma_{\rm peak}$, of the flare peaks, which corresponds to the amplitude $A$ in the \citetalias{Mendoza_2022} flare template. By using $\sigma_{\rm peak}$ to represent the amplitude, we ensure that the simulating flux is directly comparable to the standardized flux of the preliminary flare candidates, which enables us to use the standardized flux as input for the RF classifier without any additional transformation. Additionally, since \citetalias{Howard_2022} provides only the rise and decay times of the flares without their FWHM, we apply an approximate method to obtain them. Based on the \citetalias{Mendoza_2022} flare template, the rise time is roughly equal to $0.6 \times$FWHM. We therefore divide the rise time of each flare by 0.6 to estimate the corresponding FWHM for each flare.

\begin{figure}
\epsscale{1.1}
\plotone{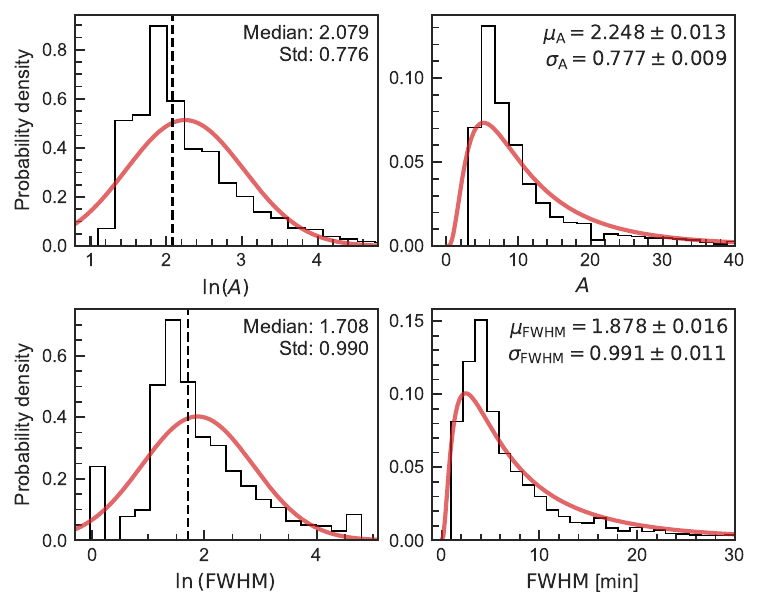}
\caption{Histograms of the flare amplitude $A$ and FWHM values, along with their natural logarithms, for the flare samples from \citetalias{Howard_2022}. Left panels: The distributions of the natural logarithms of amplitude and FWHM (black step lines) with the fitted normal distributions (red solid lines). Right panels: The distributions of the original amplitude and FWHM values (black step lines) with the fitted log-normal distributions (red solid lines). The median values (vertical dashed lines) and standard deviations of the natural logarithms of amplitude and FWHM, provided in the upper right of the left panels, are used to initialize the walkers for the MCMC analysis. The resulting estimates from the MCMC analysis are shown in the upper right of the right panels.}
\label{fig:histogram}
\end{figure}

The flare amplitude and FWHM generally exhibit a strong correlation, with higher amplitude flares tending to have longer FWHMs. In \citetalias{Howard_2022}, the histograms of these parameters follow skewed distributions with long tails toward larger values, characteristic of log-normal distributions (Figure \ref{fig:histogram}). Since taking the natural logarithm of a log-normally distributed random variable yields a normal distribution, we transform the amplitude and FWHM values by taking their natural logarithms. This allows us to simplify the fitting by using a bivariate normal distribution for the transformed data. A bivariate normal distribution is a two-dimensional generalization of the normal distribution, parameterized by the means, standard deviations, and correlation coefficient of the two variables. Specifically, we fit for the means of the natural logarithms of amplitude ($\mu_{\rm A}$) and FWHM ($\mu_{\rm FWHM}$), the standard deviations of the natural logarithms of amplitude ($\sigma_{\rm A}$) and FWHM ($\sigma_{\rm FWHM}$), and the correlation coefficient ($\rho$) between the natural logarithms of amplitude and FWHM.

We use the Python package \texttt{emcee} \citep{Foreman-Mackey_2013} to perform a Markov Chain Monte Carlo (MCMC) analysis to fit the model. The walkers are initialized by slightly perturbing the median and standard deviation of the transformed data shown in Figure \ref{fig:histogram} for the means and standard deviations of the natural logarithm of amplitude and FWHM. The initial correlation coefficient $\rho$ for each walker is randomly drawn from a uniform distribution over $[0,1]$. We run \texttt{emcee} using 128 walkers for \num{10000} steps, discarding the first 1000 steps as burn-in, which is sufficient for the chains to converge. The acceptance fraction is 0.551 and the mean auto-correlation time is 51.5 steps. The resulting estimate for $\rho$ is $0.093\pm0.016$, while the estimates for the other parameters are shown in Figure \ref{fig:histogram}.

\subsubsection{Generating Simulated Events}

Using the best-fit parameters obtained from the MCMC analysis, we define a bivariate normal distribution to generate paired samples of the natural logarithm of amplitude and FWHM values, which are then input into the \citetalias{Mendoza_2022} flare model to produce a series of simulated flare events. As the flux of these events is standardized, it is essential to introduce noise modeled by a standard normal distribution to create a more realistic representation of the observed data, ensuring that the simulation closely mimics the properties observed in intrinsic flare data.

In addition to simulating flare events, we also generate simulations of non-flare events to establish a contrasting group in the training set. To distinguish these events from flares, which exhibit an asymmetric profile characterized by a rapid rise and gradual decay, we employ a symmetric Gaussian profile parameterized by sigma and amplitude. The sigma values are derived from a log-normal distribution with parameters $\mu=3$ and $\sigma=1$, while the amplitude values are sampled from a half-normal distribution with parameter $\sigma=1$. Through this parameterization approach, the simulated flare events have a broader range of amplitude values that extend to considerably higher levels, and generally exhibit increased duration compared to non-flare events. These discrepancies in amplitude and duration concur with the intrinsic divergences between flares, which typically demonstrate substantial brightness enhancements over prolonged durations, and non-flares such as noise and artifacts, which display more subtle amplitude variations over shorter timespans. As with the simulated flare events, the simulated non-flare events are incorporated into noise modeled by a standard normal distribution. Both types of simulated events have a 2-minute cadence, which is consistent with the TESS photometry we used here.

Before incorporating the simulated events into the training set, we perform additional steps to ensure their close resemblance to the preliminary flare candidates. In accordance with the criteria used for identifying preliminary flares in Section \ref{sec:searching_candidates}, we require the generated events to have a minimum of two consecutive epochs with flux values greater than 3. Subsequently, we extend these epochs using the previously described algorithm, which involves expanding the left and right sides of each candidate event until one and two consecutive epochs, respectively, have flux values less than 1, and then truncating the left epochs. Moreover, for non-flare events with five or more epochs, we require that the peak epoch must be located on the right side of the event, thereby creating a clearer distinction between flare and non-flare events in the training set.

\begin{figure*}
\gridline{\fig{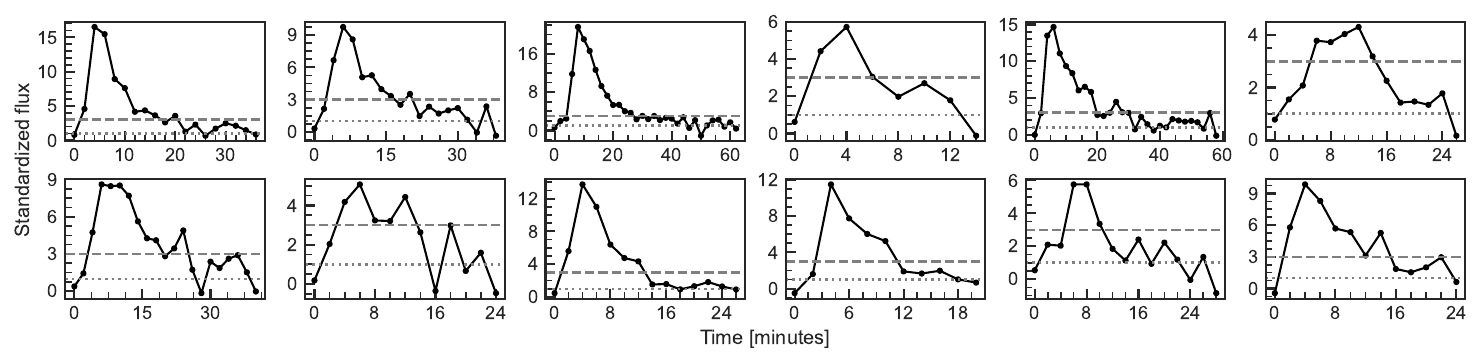}{\textwidth}{(a)}}
\vspace{-0.5cm}
\gridline{\fig{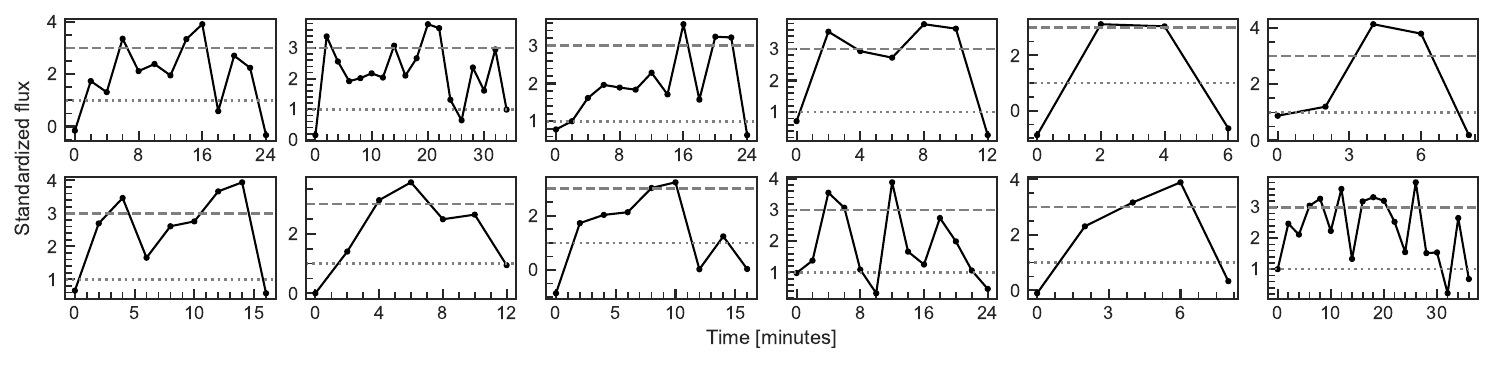}{\textwidth}{(b)}}
\caption{Examples of simulated events in the training set: (a) Flares, (b) Non-flares. The dashed lines and dotted lines indicate standardized flux values of 3 and 1, respectively.}
\label{fig:simulation}
\end{figure*}

We generate a total of 5,000 simulated events for flares and non-flares, separately. These events are subsequently partitioned into training, validation, and test sets following a 3:1:1 ratio. A selection of events from the training set is illustrated in Figure \ref{fig:simulation}.

\subsection{Feature Selection and Extraction} \label{subsec:feature}

Feature extraction is essential in machine learning applications, as it converts time-series data with varying durations into structured vectors suitable for input into the classifier. Moreover, the careful selection of an appropriate and informative feature set is also crucial to ensure the critical information embedded within the data. The chosen features should effectively encapsulate the intrinsic characteristics of the events, allowing the classifier to distinguish more accurately between distinct categories and yield more reliable results.

In this study, we employ the Python package \texttt{tsfresh} \citep{Christ_2018} to facilitate feature selection and perform feature extraction. This package is designed for machine learning applications and is capable of automatically identifying and extracting relevant features from time-series data using hypothesis testing. We initially use \texttt{tsfresh} to select a comprehensive list of features, from which we subsequently handpick a set of valuable features that efficiently distinguish between flare and non-flare events. These selected features are presented in Table \ref{tab:features}.

For the training of the RF classifier, we transform each simulated event into a single $9 \times 1$ vector, corresponding to the selected features. The feature extraction process not only vectorizes the events but also emphasizes the distinguishing characteristics that differentiate flare and non-flare events, enabling the classifier to focus on the most significant aspects of the data.

\begin{deluxetable*}{rcl}
\tablecaption{All 9 Statistical Features Selected for Classification.}
\tablehead{
\colhead{Feature} & \colhead{Importance} & \colhead{Description}
}
\startdata
abs\_energy & 0.040 & Sum over the squared values of the time series \\
first\_location\_of\_maximum & 0.316 & Relative location of the first maximum value in the time series \\
index\_mass\_quantile ($q=0.5$) & 0.131 & Mass center of the time series \\
kurtosis & 0.025 & Kurtosis of the time series \\
length & 0.018 & Length of the time series \\
maximum & 0.173 & Highest value of the time series \\
root\_mean\_square & 0.179 & Root mean square of the time series \\
skewness & 0.042 & Sample skewness of the time series \\
standard\_deviation & 0.076 & Standard deviation of the time series \\
\enddata
\tablecomments{These features are computed using the Python package \texttt{tsfresh} \citep{Christ_2018}. For details on these features and algorithms, see \url{https://tsfresh.readthedocs.io/en/latest/text/list_of_features.html}.}
\label{tab:features}
\end{deluxetable*}

\subsection{Hyperparameter Tuning and Model Training}

Following feature extraction, we proceed with hyperparameter tuning, a crucial step in optimizing the classifier's performance. This process involves determining the optimal combination of hyperparameters, which are selected based on their capacity to maximize the score of the RF model on the validation set.

We employ the Python package \texttt{optuna} \citep{Akiba_2019} to tune the following hyperparameters of the RF classifier within the specified ranges, considering a balance between model complexity and computational efficiency:
\begin{itemize}
\item \texttt{n\_estimators}: The number of trees in the forest. Range: (100, 1000, step=100)
\item \texttt{max\_depth}: The maximum depth of the trees. Range: (3, 17, step=2)
\item \texttt{max\_features}: The number of features to consider when seeking the best split. Range: (2, 5, step=1)
\end{itemize}
The \texttt{GridSampler} in \texttt{optuna} is used to perform a grid search, training the RF classifier with all possible combinations of hyperparameters and evaluating their accuracy scores on the validation set. The following hyperparameters produce the highest accuracy score (0.993) on the validation set:
\begin{itemize}
\item \texttt{n\_estimators}: 100
\item \texttt{max\_depth}: 7
\item \texttt{max\_features}: 2
\end{itemize}
We select the corresponding model as the final model. Its performance is assessed on the test set, yielding an F1 score of 0.994, which is calculated as the harmonic mean of precision and recall.

Moreover, we can extract the feature importance after the training process, which assigns an importance score to each feature based on its contribution to the decision-making process. The significance of our selected features is listed in Table \ref{tab:features}, which represents the relative contribution of each feature to the classification process, offering insights into their relevance in the context of flare identification. The \texttt{first\_location\_of\_maximum} feature, capturing the relative location of peak flux, emerges as the most important, which is reasonable given that flares exhibit a rapid rise and gradual decay.

\subsection{Validating Preliminary Flare Candidates}

After training the RF classifier, we apply it to the preliminary flare candidates. As outlined in Section \ref{subsec:feature}, we extract the corresponding set of features for these candidates. In cases where the candidates contain missing values, we employ a linear fitting method to interpolate them prior to feature extraction. The resulting feature vectors are then fed as input into the RF classifier.

The RF classifier computes a probability score for each candidate, indicative of its likelihood of being an intrinsic flare. We filter the preliminary flare candidates by applying a probability threshold of 0.5. Candidates with a probability score above this threshold are classified as validated flare events, while others are considered false positives. This filtering effectively refines the flare candidate list, retaining only the events with a high probability of being intrinsic flares for further investigation.

To further improve the reliability and accuracy of the flare validation process, we perform an additional manual verification of the filtered flare candidates through visual inspection. During this process, we identify several common sources of false positives, including glitches in the light curves, unclassified CVs in the SIMBAD database, SSO encounters not captured in the SkyBoT query (potentially due to unknown SSOs or bright SSOs passing by without falling within the queried cone area), and low amplitude candidates exhibiting flare-like profiles but suffering from poor light curve quality.

\section{Results} \label{sec:results}

From a total of \num{38102} available 2-minute cadence light curves of \num{11618} compact stars, we identified 7584 events as preliminary flare candidates. After excluding 578 events that are attributed by SSOs, we validated the remaining 7006 events with the help of our RF classifier, which flagged 3558 of these events as false positives. Finally, we confirmed 1016 flares to be real events with further visual inspection of the remaining 3448 candidates. These confirmed flare events originated from 193 compact targets, which include 182 events from 58 sdB/sdO stars and 834 events from 135 WDs. Table \ref{tab:flare} provides the observed properties of all confirmed flare events, such as peak time, amplitude, and energy in the TESS bandpass (see Section \ref{sec:flare_energy}). Table \ref{tab:flare_star} presents the key attributes of the flaring compact stars, including their classification, flare occurrence frequency, and logarithmic fractional flare luminosity (see Section \ref{sec:frac_flare_lum}).

\begin{deluxetable*}{rcccccccC}
\setlength{\tabcolsep}{8pt}
\tablecaption{Catalog of All 1016 Flares Observed across \num{38102} Light Curves of \num{11618} Compact Stars at \qty{120}{\second} Cadence during TESS Cycles 1-5.\label{tab:flare}}
\tablehead{
\colhead{TIC} & \colhead{Sector} & \dcolhead{t_{\rm start}} & \dcolhead{t_{\rm peak}} & \dcolhead{t_{\rm stop}} & \colhead{SNR} & \dcolhead{A} & \colhead{ED} & \dcolhead{E_{\rm TESS}} \\
\colhead{} & \colhead{} & \colhead{(BTJD)} & \colhead{(BTJD)} & \colhead{(BTJD)} & \colhead{} & \dcolhead{(\Delta F/F)} & \colhead{(s)} & \colhead{(erg)}
}
\startdata
6997163 & 46 & 2576.0223 & 2576.0251 & 2576.0446 & 4.65 & 0.101 & 110.3(9.5) & 1.03(0.20)10^{35} \\
20656977 & 20 & 1845.9203 & 1845.9230 & 1845.9439 & 13.17 & 0.577 & 414.9(24.2) & 1.53(0.28)10^{33} \\
21860382 & 52 & 2719.4594 & 2719.4622 & 2719.4844 & 4.07 & 1.183 & 1097.5(99.7) & 1.05(0.10)10^{33} \\
23226265 & 9 & 1546.4458 & 1546.4472 & 1546.4597 & 6.51 & 0.188 & 122.0(11.0) & 1.20(0.26)10^{32} \\
23226265 & 9 & 1547.6653 & 1547.6694 & 1547.6750 & 3.69 & 0.106 & 54.4(8.6) & 5.37(1.14)10^{31} \\
23226265 & 9 & 1559.3961 & 1559.4002 & 1559.4113 & 4.95 & 0.146 & 94.9(9.9) & 9.36(1.99)10^{31} \\
23226265 & 9 & 1566.0406 & 1566.0600 & 1566.0892 & 10.76 & 0.318 & 538.0(20.1) & 5.30(1.13)10^{32} \\
23226265 & 36 & 2283.5776 & 2283.5818 & 2283.5943 & 14.51 & 0.294 & 109.6(6.1) & 1.08(0.18)10^{32} \\
\enddata
\tablecomments{
The table presents the details of the 1016 flare events. Columns are TIC ID, Sector, start time, peak time, stop time, signal-to-noise ratio, amplitude, equivalent duration, and energy in the TESS bandpass. The start, peak, and stop time are in Barycentric TESS Julian Date (BTJD). The signal-to-noise ratio (SNR) is the maximum value in the standardized light curve during the flare event. Uncertainties are given in parentheses.\\
(This table is available in its entirety in machine-readable form.)
}
\end{deluxetable*}

\begin{deluxetable*}{rccCccCc}
\setlength{\tabcolsep}{5pt}
\tablecaption{Catalog of All 193 Flaring Compact Stars.\label{tab:flare_star}}
\tablehead{
\colhead{TIC} & \colhead{Spectral Type} & \colhead{Object Type} & \dcolhead{L_{\rm TESS}} & \dcolhead{N_{\rm flares}} & \colhead{Flare freq.} & \dcolhead{\log(L_{\rm flare}/L_{\rm TESS})} & \colhead{Poll.} \\
\colhead{} & \colhead{} & \colhead{} & \colhead{(erg/s)} & \colhead{} & \dcolhead{({\rm Sector}^{-1})} & \colhead{} & \colhead{}
}
\startdata
6997163 & sdB+F/G/K & HotSubdwarf & 9.37(1.06)10^{32} & 1 & 0.5 & -4.54 & 0 \\
20656977 & DA+M & WhiteDwarf & 3.69(0.04)10^{30} & 1 & 1.0 & -3.68 & 0 \\
21860382 & DA & WhiteDwarf & 9.54(0.07)10^{29} & 1 & 0.333 & -3.60 & 2 \\
23226265 & cand-DAV & WhiteDwarf & 9.86(0.06)10^{29} & 6 & 3.0 & -3.58 & 1 \\
23385704 & WD & WhiteDwarf\_Candidate & 1.23(0.01)10^{30} & 1 & 0.333 & -4.20 & 2 \\
23992223 & DA & WhiteDwarf & 6.20(0.05)10^{30} & 1 & 0.5 & -4.78 & 1 \\
\enddata
\tablecomments{The table presents the details of the 193 flaring compact stars. Columns are TIC ID, spectral type given by TASC~WG8 target list, object type queried from the SIMBAD database, stellar luminosity in the TESS bandpass, number of observed flares, flare occurrence frequency per TESS sector, logarithmic fractional flare luminosity, and pollution level (Poll.) of the target (see Section \ref{sec:pollution_level}). The description of the object types is presented in \url{https://simbad.cds.unistra.fr/guide/otypes.htx}. Uncertainties are given in parentheses.\\
(This table is available in its entirety in machine-readable form.)
}
\end{deluxetable*}

\subsection{Flare Properties and Fractional Flare Luminosity}

\begin{figure*}
\gridline{\fig{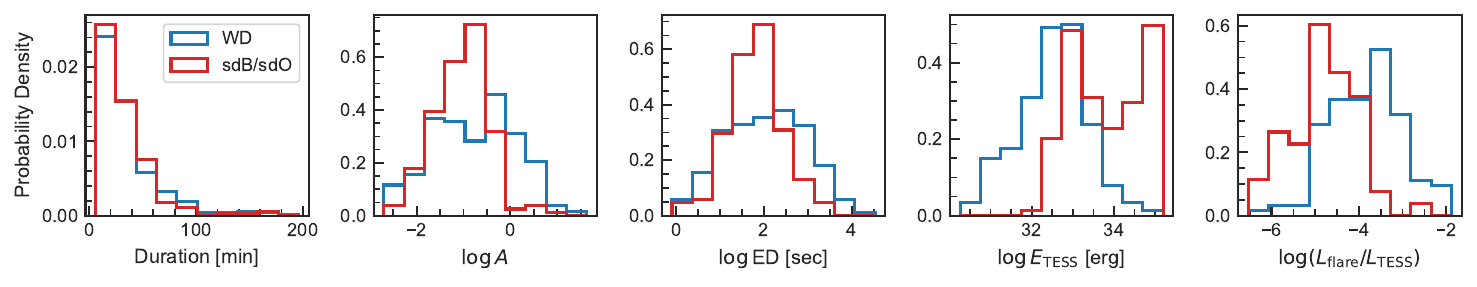}{\textwidth}{(a)}}
\vspace{-0.25cm}
\gridline{\fig{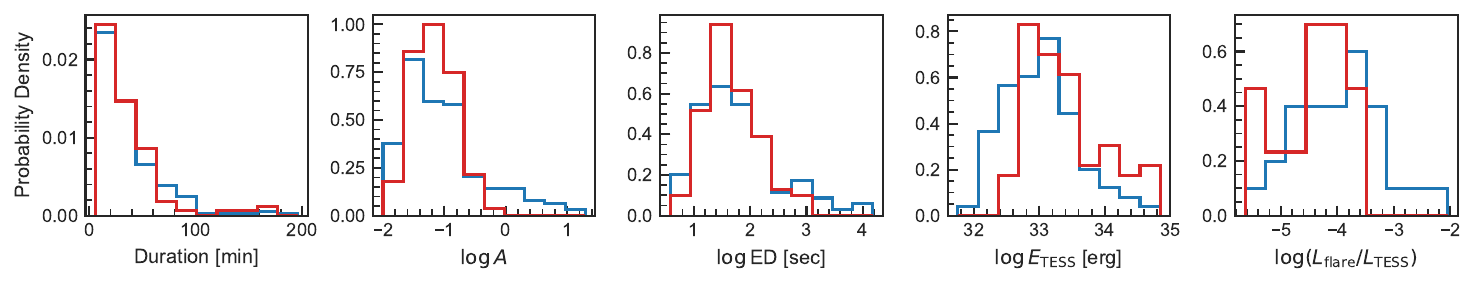}{\textwidth}{(b)}}
\caption{Histograms comparing flare properties and fractional flare luminosity between WDs (blue lines) and sdB/sdO stars (red lines). The top panels (a) show distributions for all flare events and flaring compact stars, while the bottom panels (b) display only flare events and flaring compact stars with a PL of 0. The left four panels in both rows show flare property distributions: duration, logarithmic amplitude, logarithmic ED, and logarithmic energy in the TESS bandpass, either for (a) 834 WD and 182 sdB/sdO flares or (b) 193 WD and 86 sdB/sdO flares. The right panels illustrate distribution of the logarithmic fractional flare luminosity value, either for (a) 135 flaring WDs and 58 flaring sdB/sdO stars or (b) 28 flaring WDs and 16 flaring sdB/sdO stars.}
\label{fig:stats}
\end{figure*}

For each identified flare event, we measure properties including duration, amplitude, equivalent duration, and energy in the TESS bandpass. We also compute the fractional flare luminosity for each of the flaring compact stars to quantitatively measure their strength of flare activity. Figure \ref{fig:stats}(a) shows histograms comparing the distributions of flare properties and fractional flare luminosity between WDs and sdB/sdO stars, while comparisons exclusively between compact stars with a pollution level of 0 (see Section \ref{sec:pollution_level}) are shown in Figure \ref{fig:stats}(b).

\subsubsection{Flare Amplitude and Equivalent Duration}

The flare amplitude, defined as the maximum increase in flux during the flare event relative to the flux of the star in its quiescent state, is calculated as:
\begin{equation}
A = \frac{F_{\rm peak} - F_{\rm quiescent}}{F_{\rm quiescent}},
\end{equation}
where $A$ represents the flare amplitude, $F_{\rm peak}$ is the maximum flux observed during the flare event and $F_{\rm quiescent}$ is the quiescent flux.

The equivalent duration \citep[ED;][]{Gershberg_1972} provides an estimate of the energy output of a flare relative to the quiescent emission of the star. Expressed in units of time (e.g., seconds), the ED represents the duration for which the star would need to emit at its quiescent brightness level to equal the excess energy released by the flare. It is calculated as follows,
\begin{equation}
{\rm ED} = \int \frac{F_{\rm flare}(t) - F_{\rm quiescent}}{F_{\rm quiescent}} {\rm d}t,
\end{equation}
where the integral is computed by a trapezoidal sum of the light curve between the start and stop times of the flare. The uncertainty of ED is calculated following \citet{Davenport_2016}.

\subsubsection{Flare Energy in the TESS Bandpass} \label{sec:flare_energy}

Typically, the total energy emitted by a stellar flare (i.e., the bolometric flare energy) can be estimated if the effective temperature and radius of the star are known \citep[see, e.g.,][]{Shibayama_2013, Gunther_2020}. However, the result of the cross-matching of the flaring compact stars with Gaia DR3 shows that only a small subset (\num{\sim10}\%) of our samples have these stellar parameters available, which restricts our ability to calculate the bolometric flare energy across the entire sample.

Given these limitations, we refocused towards estimating the flare energy in the TESS bandpass ($E_{\rm TESS}$), a measure that provides insight into the observable energy released during the flare event. It is noteworthy that the bolometric flare energy can be significantly greater, by a factor of approximately five or more, than the flare energy estimated in the TESS bandpass \citepalias[see][]{Howard_2022}. This discrepancy arises because flares, often approximated by a \qty{9000}{\K} blackbody \citep{Jackman_2023}, primarily emit energy in the high-energy range at short wavelengths, which fall outside the TESS bandpass.

We first calculate the luminosity of the star in the TESS bandpass using the following formula:
\begin{equation}
L_{\rm TESS} = 4\pi d^2 \times F_{\rm TESS},
\end{equation}
where $d$ is the distance to the star derived from its parallax listed in Gaia DR3. The flux of the star in the TESS bandpass $F_{\rm TESS}$ is calculated via the following relation:
\begin{equation}
F_{\rm TESS} = 10^{-0.4 T} \times F_0,
\end{equation}
where $T$ is the TESS magnitude derived from the TESS Input Catalog version 8.2 \citep[TIC v8.2;][]{Stassun_2019, Paegert_2021} and $F_0=4.03 \times 10^{-6}~{\rm erg\,s^{-1}\,cm^{-2}}$, which is the flux corresponding to $T=0$ \citep{Sullivan_2015}. The uncertainty in the calculated luminosity $L_{\rm TESS}$ accounts for uncertainties in both the TESS magnitude and the parallax.

Lastly, we compute the flare energy in the TESS bandpass by:
\begin{equation}
E_{\rm TESS} = {\rm ED} \times L_{\rm TESS}, \label{eqn:etess}
\end{equation}
where ED is the equivalent duration of the flare and $L_{\rm TESS}$ is the calculated luminosity of the star in the TESS bandpass.

For seven of the flaring compact stars, which correspond to 121 flare events, the parallax parameters were not available in Gaia DR3. As a result, we were only able to calculate the energies for the remaining 923 flares. The median value of uncertainties associated with $E_{\rm TESS}$ is 22.1\%.

\subsubsection{Fractional Flare Luminosity} \label{sec:frac_flare_lum}

For each flaring compact star, we calculate the fractional flare luminosity, represented as $L_{\rm flare}/L_{\rm TESS}$, which is the total luminosity emitted by flares relative to that emitted by the star through the TESS bandpass. Serving as an intuitive metric, it quantifies the intensity of stellar flare activity. The metric was originally denoted as $L_{\rm fl}/L_{\rm Kp}$ in \citet{Lurie_2015} for characterizing the flare activity level of the flaring stars with Kepler photometry. This metric is now widely used in various flare research \citep[see, e.g.,][]{Davenport_2016, Davenport_2019}. In certain cases, the $L_{\rm Kp}$ has been substituted with the bolometric luminosity $L_{\rm bol}$ \citep[see, e.g.,][]{Yang_2019, Ilin_2021}.

We calculate this ratio for a star using the following equation,
\begin{equation}
\frac{L_{\rm flare}}{L_{\rm TESS}} = \frac{\sum {\rm ED}_i}{t_{\rm obs}},
\end{equation}
where ${\rm ED}_i$ represents the ED of each flare for the star, and $t_{\rm obs}$ is the total observation time for the star by TESS. A higher value of this ratio signifies enhanced flare activity, indicative of an increased level of stellar magnetic activity.

\begin{figure}
\epsscale{1.2}
\plotone{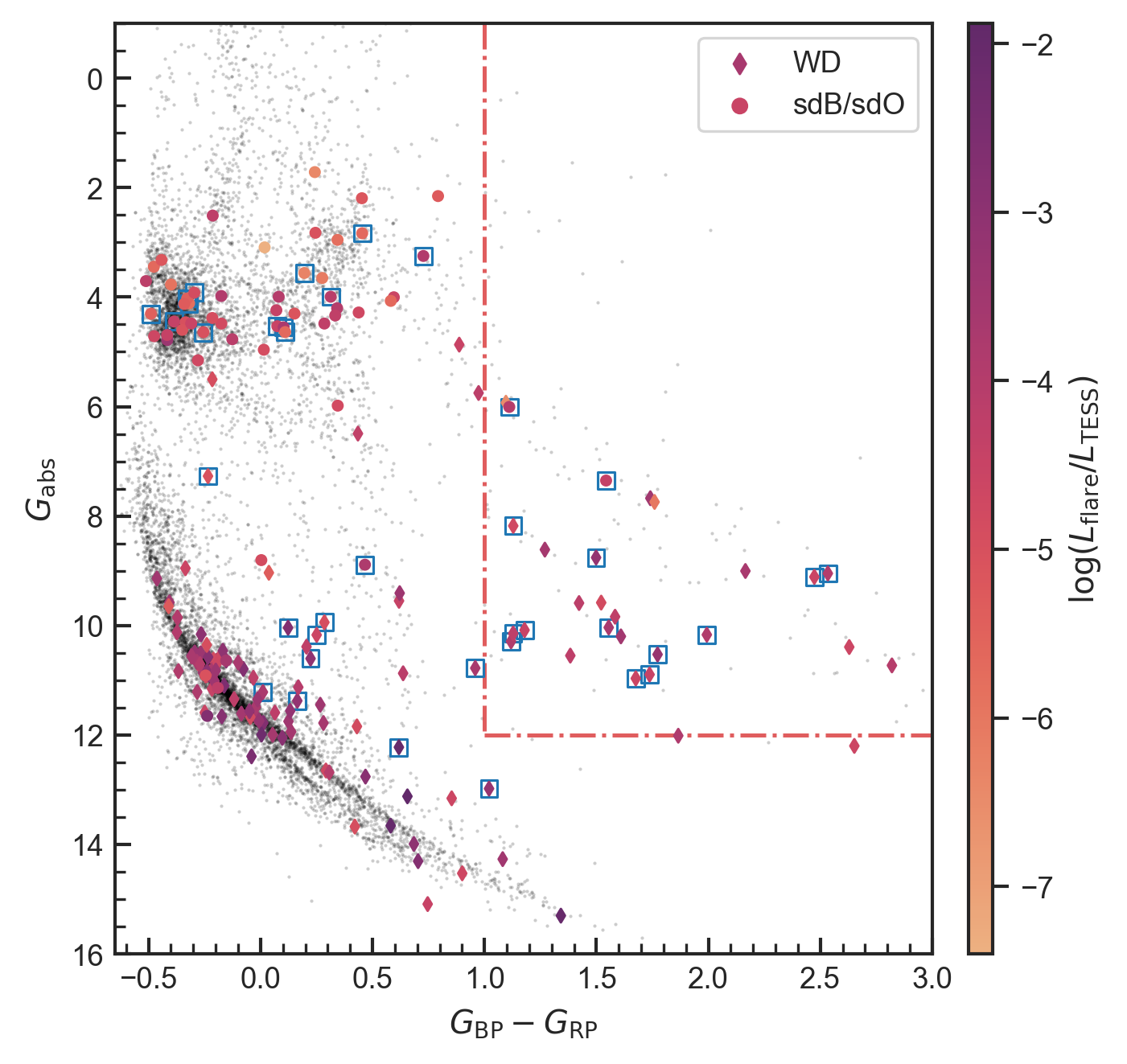}
\caption{Gaia H--R diagram of the flaring compact stars included in our study. The flaring WDs and sdB/sdO stars are signified by diamond and circle markers, respectively, with other compact stars from TASC~WG8 target list denoted with black dots. The colors of the markers correspond to the $\log (L_{\rm flare}/L_{\rm TESS})$ value for each flaring star. The stars surrounded by blue squares are those with a pollution level of 0 (see Section \ref{sec:pollution_level}). The red dash-dotted lines indicate the thresholds used for filtering out stars with companions (see Section \ref{sec:companion_stars}).}
\label{fig:H--R}
\end{figure}

We cross-matched the flaring compact stars with Gaia Data Release 3 \citep[DR3;][]{Gaia_2016, Gaia_2023} to acquire their BP-RP color index ($G_{\rm BP} - G_{\rm RP}$) and absolute Gaia G magnitude ($G_{\rm abs}$), and plot them on the Gaia Hertzsprung--Russell (H--R) diagram (Figure \ref{fig:H--R}). We distinguish between sdB/sdO stars and WDs in the diagram and illustrate the fractional flare luminosity across the flaring compact stars.

\subsection{Contamination Check}

Prior to analyzing the flare activities, it is vital to refine our sample to more accurately characterize the flare activity inherent to sdB/sdO stars and WDs. This refinement ensures the exclusion of targets where flare activities might potentially originate from contaminating sources other than the compact stars themselves, such as nearby objects or companion stars of the compact target.

\subsubsection{Evaluating Pollution Level} \label{sec:pollution_level}

\begin{figure*}
\epsscale{1.2}
\gridline{\fig{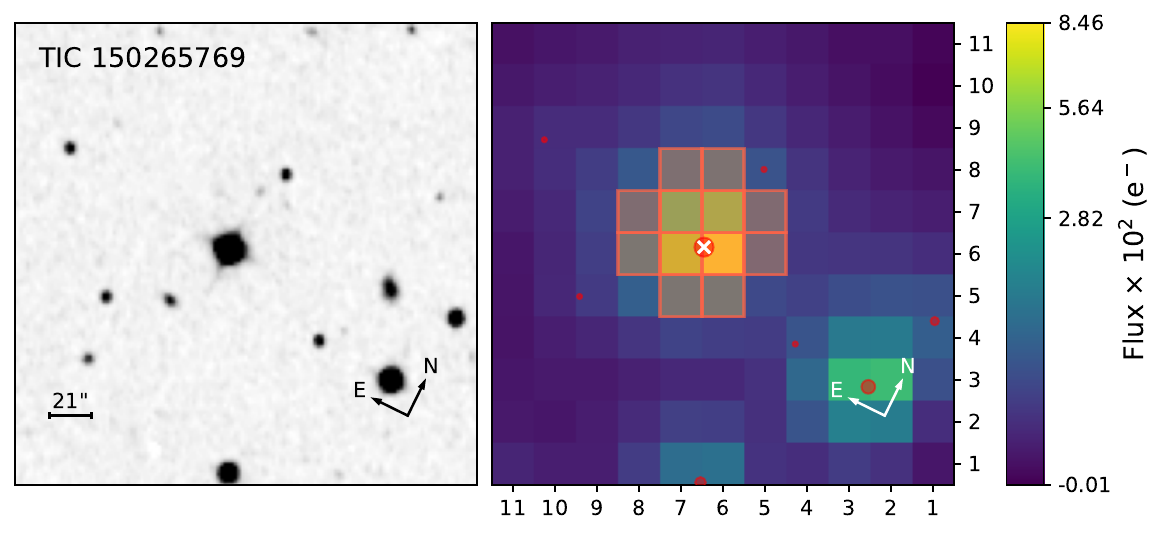}{\columnwidth}{(a) PL=0} \hspace{-1cm}
          \fig{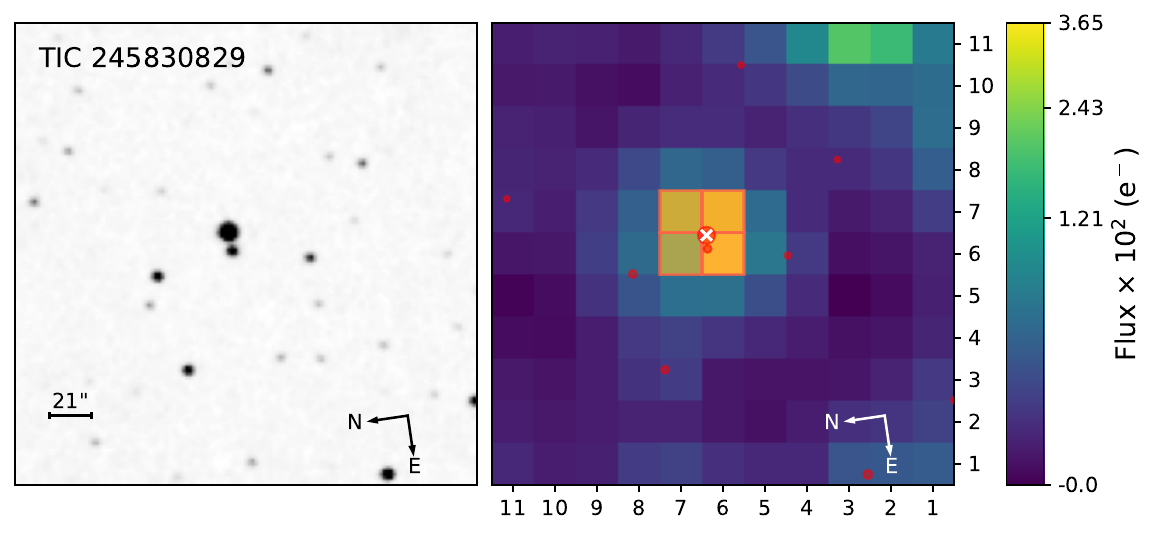}{\columnwidth}{(b) PL=1}}
\gridline{\fig{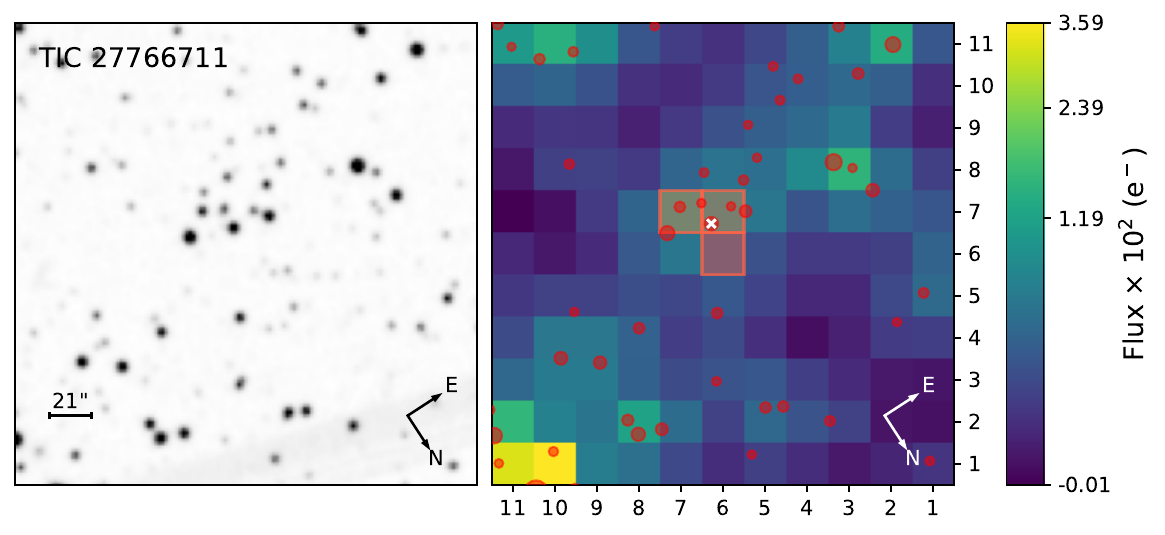}{\columnwidth}{(c) PL=2} \hspace{-1cm}
          \fig{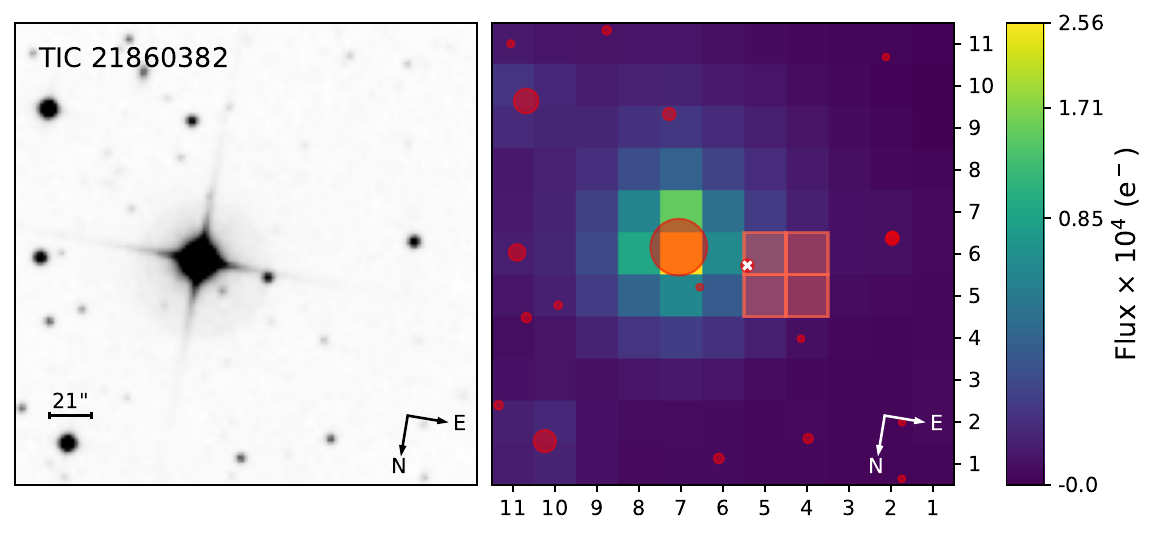}{\columnwidth}{(d) PL=2}}
\caption{Examples of the identification charts provided by \texttt{tpfi}: (a): an ideal situation that the aperture only hosts the target with no other bright stars nearby, (b): the aperture hosting the target star and another dimmer star, (c): the aperture hosting numerous stars with comparable brightness alongside the target, (d): only the target resides within the aperture, but in proximity to a bright star. In each chart, the right panel overlays the Gaia DR3 catalog onto the target pixel file, with the target denoted by a white cross symbol. The size of the circle represents the relative brightness of the stars, as indicated by the Gaia G magnitude. The red region indicates the default aperture mask used by SPOC for photometry extraction. The left panel shows the same sky coverage, using DSS2 red images, with the same orientation.}
\label{fig:tpfi}
\end{figure*}

Due to the relatively low angular resolution of TESS (\qty{21}{arcsec.pixel^{-1}}), there is a considerable risk that the photometry may be contaminated by nearby objects. To address this concern, we developed an open-source script, \texttt{tpfi}, which generates identification charts for TESS target pixel files, as illustrated in Figure \ref{fig:tpfi}. Our script, based on \texttt{tpfplotter}\footnote{\url{https://github.com/jlillo/tpfplotter}} \citep{Aller_2020}, introduces a notable feature: the ability to visualize the identical sky coverage of the target pixel file as provided by the Digital Sky Survey (DSS)\footnote{\url{https://archive.eso.org/dss/dss}}. This enhancement allows for a more convenient assessment of the contamination level of a target. In addition to its applicability to TESS, we extended the use of \texttt{tpfi} to Kepler and K2 missions. We anticipate that this enhancement will make the script valuable to broader communities, for instance, the exoplanet and variable star research communities. Our script is publicly available on Github\footnote{\url{https://github.com/keyuxing/tpfi}}.

Following these considerations, we employed \texttt{tpfi} to create identification charts for all compact stars in our sample. We thus can conduct a visual examination of each of the 193 compact stars in our sample to assess the potential for contamination from nearby sources. In order to systematically quantify the contamination, we defined the pollution level (PL) for each star, as shown in Table \ref{tab:flare_star}. A PL=0 indicates that the target is the only object within the aperture, without any significant polluting photometry from nearby stars. A PL=1 denotes that the target is the brightest object within the aperture, but there are other dim stars present. Lastly, a PL=2 refers to the cases where the target is not the brightest object within the aperture or near a much brighter star. To illustrate, Figure \ref{fig:tpfi}(a) represents a scenario with a PL=0, indicating an unpolluted target star. Figure \ref{fig:tpfi}(b) displays minor contamination, warranting a PL=1. Cases of severe contamination, as demonstrated by Figures \ref{fig:tpfi}(c) and (d), are assigned with a PL=2.

In our categorization, 44 stars have PL=0, 62 stars fall into the PL=1 category, and 87 stars are classified with PL=2, which corresponds to 279, 256, and 481 flares, respectively. These pollution levels offer a crude estimate of potential contamination. They are particularly useful in downstream analyses, where the potential impact of such contamination on our flare detection results must be considered. We also note that a high PL does not necessarily disqualify a star as a flare candidate. However, stars with a high PL warrant additional caution and follow-up investigation to confirm the source of the flare events.

\subsubsection{Detecting companion stars} \label{sec:companion_stars}

To address potential contamination from companion stars, we firstly exclude flaring stars that are labelled as binary in TASC~WG8 target list. We then locate the rest flaring stars on the Gaia H--R diagram. If a target falls on the main sequence, it indicates the compact star has a brighter MS companion dominating the observed brightness, which is especially relevant for the intrinsically faint WDs. Therefore, we exclude the flaring compact stars positioned within the main sequence of the Gaia H--R diagram ($G_{\rm BP} - G_{\rm RP}> 1$ and $G_{\rm abs} < 12$, see Figure \ref{fig:H--R}). We also exclude the stars that are not provided with $G_{\rm BP} - G_{\rm RP}$ or $G_{\rm abs}$ in Gaia DR3, as their positions on the H--R diagram are unavailable. After the filtering, 13 flaring compact stars remained, comprising seven WDs and six sdB/sdO stars.

To further examine the likelihood of companion stars, we constructed the spectral energy distributions (SEDs) for these stars using the VO Sed Analyzer (VOSA)\footnote{\url{http://svo2.cab.inta-csic.es/theory/vosa}} \citep{Bayo_2008}. Our examination revealed that all seven WDs display a significant red/near-infrared flux excess, suggesting the presence of cool companions. In contrast, the SEDs of the six sdB/sdO stars displayed less noticeable red/near-infrared excess. This may be attributed to the inherent high brightness of sdB/sdO stars, which can dominate the SED when paired with a cool MS star. We then employed \texttt{speedyfit}\footnote{\url{https://github.com/vosjo/speedyfit}} to fit the SEDs of the sdB/sdO stars. Initially, we used spectral models exclusively from the Tübingen NLTE Model-Atmosphere Package \citep[TMAP;][]{Werner_1999, Werner_2003, Rauch_2003}. However, the fitting residuals indicated a clear IR excess beyond what the model predicted for all six sdB/sdO stars. This discrepancy was resolved when we incorporated a second component in the fitting, modeled using ATLAS9 spectra \citep{Castelli_2003}. The results of the MCMC fitting suggest that the effective temperatures of the cool companions are within the range of \qtyrange{3500}{4500}{\K}, consistent with K/M-dwarfs. A more detailed description of the fitting process of \texttt{speedyfit} is given in \citet{Vos_2017, Vos_2018}.

\subsubsection{Selecting the Refined Sample} \label{sec:refined_sample}

Our previous analysis found that all flaring compact stars show evidence of contamination from nearby objects or the presence of an unresolved companion. As a result, we have been unable to conclusively attribute any specific stellar flares solely to an individual compact star. However, the origin of the stellar flares in binary systems containing compact stars is still unclear and demands careful examination.

For these flaring compact stars with cool companions, the origin of the flares warrants careful consideration, since flares are common in cool MS stars, particularly M-dwarfs. Therefore, for WDs, there is a high probability that the flares originate from their red companions, especially given the comparable luminosity levels of WDs and cool MS stars. This makes it reasonable for flares from the cool companions to be detectable in the combined light curve of the WD binary systems. However, the situation is less clear-cut for sdB/sdO systems. The comparative high brightness of sdB/sdO stars over cool MS stars suggests the flare contribution from a cooler companion would be less apparent. If the stellar flares indeed originate from the cool companion, they would need to be extraordinarily energetic to be noticeable in the combined light curve. Such high-energy flares from a cool MS companion are relatively rare, implying that the flares observed in sdB/sdO systems may have a different origin. Another possible explanation is that the companions to compact objects are more magnetically active compared to their isolated counterparts, causing more frequent high-energy flares.

To conduct a more detailed investigation into the origin of flares in sdB/sdO systems, a refined sample of sdB/sdO stars is necessary for further in-depth analysis of flare activity. To ensure the purity of the refined sample, we cross-match all 16 flaring sdB/sdO stars with PL=0 with the known hot subdwarf catalog in \citet{Culpan_2022}. There are three stars not present in the known hot subdwarf catalog, which are misclassified hot subdwarfs or hot subdwarf candidates, but labelled as hot subdwarf in TASC~WG8 target list. This left us with a refined sample of 13 sdB/sdO stars, corresponding to 23 flare events. The detailed parameters of the 13 selected sdB/sdO stars are presented in Table \ref{tab:flare_star_refined}, while the 23 flare events are shown in Fig. \ref{fig:refined_flares}.

\begin{rotatetable*}
\begin{deluxetable*}{rrcccCcccC}
\tablecaption{Catalog of Flaring Compact Stars in the Refined Sample.\label{tab:flare_star_refined}}
\tablehead{
\colhead{TIC} & \colhead{Gaia DR3 Source ID} & \colhead{Name} & \colhead{SpClass} & \dcolhead{T} & \dcolhead{G_{\rm BP} - G_{\rm RP}} & \dcolhead{G_{\rm abs}} & \dcolhead{N_{\rm flares}} & \colhead{Flare freq.} & \dcolhead{\log(L_{\rm flare}/L_{\rm TESS})} \\
\colhead{} & \colhead{} & \colhead{} & \colhead{} & \colhead{(mag)} & \colhead{(mag)} & \colhead{(mag)} & \colhead{} & \dcolhead{{\rm Sector}^{-1})} & \colhead{}
}
\startdata
6997163 & 650908345319118720 & GALEXJ08259+1307 & sdB & 14.203 & 0.0751 & 4.5317 & 1 & 0.5 & -4.54 \\
52078744 & 4704482467645621376 & GALEXJ01077-6707 & sdB & 13.861 & -0.2549 & 4.6422 & 1 & 0.167 & -5.02 \\
118327563 & 5000760581717433088 & CD-38222 & sdB & 10.515 & -0.319 & 4.1129 & 1 & 0.333 & -5.84 \\
150265769 & 781164326766404736 & Feige34 & sdO & 11.5152 & -0.4893 & 4.305 & 1 & 0.5 & -5.47 \\
167976324 & 5465148904077059968 & GALEXJ10078-2924 & sdO & 12.779 & -0.3275 & 4.0243 & 1 & 0.25 & -5.48 \\
202507151 & 1615596448547786496 & PG1524+611 & sdB+G0V & 12.39 & 0.1957 & 3.5597 & 1 & 0.143 & -6.27 \\
206688085 & 6602969028791558016 & GALEXJ22568-3308 & sdB+F & 13.269 & 0.4541 & 2.8338 & 1 & 0.25 & -5.67 \\
219974863 & 6468929937072637312 & EC20217-5704 & sdOB & 12.4 & -0.2958 & 3.9193 & 1 & 0.333 & -4.84 \\
231712886 & 6499440216512468096 & EC23257-5443 & sdB+F & 13.568 & 1.1109 & 6.0068 & 3 & 1.0 & -4.02 \\
234281664 & 6404530097924835968 & BPSCS22956-94 & He-sdB & 12.901 & 0.1036 & 4.5531 & 6 & 1.2 & -3.76 \\
262846506 & 2560685069816099968 & PHL1079 & sdB+G7V & 12.901 & 0.1101 & 4.6389 & 1 & 0.333 & -5.69 \\
368628965 & 23330399790876032 & PG0232+095 & sdB+G1V & 11.77 & 0.7273 & 3.2486 & 4 & 2.0 & -3.95 \\
443619867 & 3476266612927121536 & EC12219-2618 & sdB+MS & 14.235 & 0.3135 & 3.9936 & 1 & 1.0 & -4.12
\enddata
\tablecomments{This table lists the 13 sdB/sdO stars in our refined sample. Columns are TIC ID, Gaia DR3 source ID, target name, spectral classification given by \citet{Culpan_2022}, TESS magnitude from TIC v8.2 \citep{Stassun_2019, Paegert_2021}, BP-RP color index from Gaia DR3, absolute Gaia G magnitude from Gaia DR3, number of observed flares, flare occurrence frequency per TESS sector, and logarithmic fractional flare luminosity.}
\end{deluxetable*}
\end{rotatetable*}

\subsection{Flare Frequency Distribution}

We here investigate the Flare Frequency Distributions (FFDs) of different subsets of the flaring compact stars. The FFD, typically described by a power law \citep{Lacy_1976}, represents the frequency of flare occurrences as a function of their energy. It can be expressed as,
\begin{equation}
{\rm d}N(E) = k E^{-\alpha} {\rm d}E, \label{eqa:ffd}
\end{equation}
where $N$ represents the number of flares occurring within a specific observation duration, $E$ denotes the flare energy, $k$ is a proportionality constant, and $\alpha$ is the power-law index \citep{Jackman_2021}. This power-law index, $\alpha$, plays a critical role in investigating flare production mechanisms, and its derivation is essential to our study of flare activities in compact stars.

Due to differences in observation durations for various flaring compact stars, a direct computation of their FFD is not feasible. We hence employ the cumulative FFD to estimate $\alpha$ instead. The cumulative FFD is derived by integrating Equation (\ref{eqa:ffd}), leading to
\begin{equation}
\log(\nu) = \beta + (1-\alpha) \log(E),
\end{equation}
where $\beta=\log(\frac{k}{1-\alpha})$, and $\nu$ denotes the frequency of flares, i.e. the number of flares per unit time with energy exceeding a specific threshold.

We compute three distinct cumulative FFDs, each representing flare events from all observed sdB/sdO stars, WDs, and the refined sample of sdB/sdO stars (see Section \ref{sec:refined_sample}). For every flare event with energy $E$, we calculate the corresponding $\nu$ as follows:
\begin{equation}
\nu = \sum \frac{N_i(>E)}{t_i},
\end{equation}
where $N_i(>E)$ is the number of flares with energies greater than $E$ and $t_i$ is the observation time for each flaring compact star.

\begin{figure}
\epsscale{1.2}
\plotone{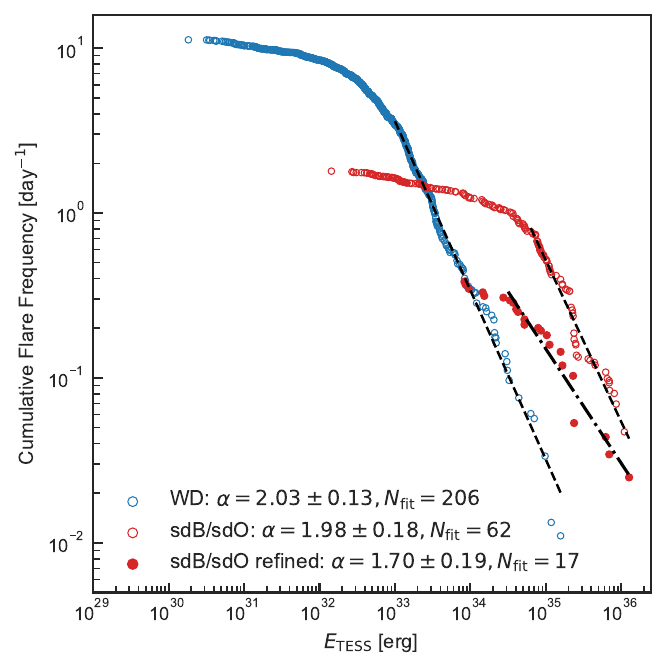}
\caption{Cumulative FFDs and corresponding power law fits. The blue and red open circle markers represent flares from all WDs and sdB/sdO stars, respectively. The red filled circle markers signify flares from the refined sdB/sdO sample. The dashed and dash-dot black lines represent the best-fit power laws to the cumulative FFDs of the full and refined samples, respectively. The fitted $\alpha$ values and uncertainties, along with the number of flares used for fitting $N_{\rm fit}$, are shown in the bottom left.}
\label{fig:cffd}
\end{figure}

Subsequently, we use the MCMC method to fit a linear regression model to each cumulative FFD. This approach effectively mitigates binning effects, providing an advantage over the conventional method of histogram generation and straight-line fitting \citep{Maschberger_2009}. We only fit the portion of the cumulative FFDs where we consider the sample complete. Figure \ref{fig:cffd} displays the cumulative FFDs with corresponding best-fit lines, the fitted $\alpha$ values and uncertainties, and the number of flares included in the fitting.

\section{Summary and Discussion} \label{sec:summary}

Based on Cycles 1-5 of TESS photometry, we comprehensively investigated the flaring activity observed in sdB/sdO stars and WDs, which identified 1016 flare events from 193 compact stars (Table \ref{tab:flare} \& \ref{tab:flare_star}). We pioneered a new method for flare detection, specifically designed to address short-term periodic variations in light curves. This method enhances flare detection capabilities when applied to the light curves, despite the complexities introduced by common phenomena in compact stars such as pulsation and binary effect. We also considered potential interferences from CVs and SSOs and eliminated them. In addition, we implemented a validation step using machine learning to effectively filter out false positives. These rigorous detection and validation processes assured the reliability of the detected flares, enabling us to establish the first flare catalog of compact stars, which contains 58 sdB/sdO stars and 135 WDs with 182 and 834 flare events, respectively.

We then thoroughly examined potential contamination caused by nearby objects and companion stars of the compact stars. This included developing an open-source script, \texttt{tpfi}, to generate identification charts for TESS target pixel files, thereby enabling a detailed visual inspection for potential contamination sources near the target. We also conducted analysis using Gaia DR3 data to identify binary systems with a bright MS companion, as well as SED fitting to reveal infrared excess from cool companion stars. Through these extensive analyses, we found evidence that all flaring compact stars showed signs of contamination from nearby objects or unresolved companion stars. As a result, we cannot conclusively attribute any specific stellar flares solely to an individual compact star at this stage.

For WDs, there is a high probability the observed flares originate from the cool MS companion. With comparable luminosities and vigorous magnetic activity generating frequent energetic flares, cool companions (e.g. K/M dwarfs) can overwhelm the emission of the WD. Considering the large number of 7505 WDs we searched for stellar flares, we still did not confidently detect flares intrinsic to any individual WD. This may partly be attributed to the low angular resolution of TESS causing severe pollution in the light curves (only $\sim25\%$ flaring compact stars have a PL=0). However, our detection provides some indication that the likelihood of observable flares occurring on WDs may be low. Alternatively, if stellar flares do occur on WDs, they may operate via a different mechanism with shorter timescales on the order of the dynamical timescale of WDs ($\sim$seconds), which cannot be captured by TESS cadence.

Nevertheless, the origin of stellar flares in sdB/sdO+MS binaries still warrants investigation. With significantly greater luminosity, a flare from the cool MS companion would need to be extraordinarily energetic to be detectable in the combined light curve from the system, which is very rare. This suggests the observed flares may be not from the cool MS companions. We aim to explore this scenario further through analyzing our refined sample of sdB/sdO stars. By focusing on high-purity sdB/sdO stars with negligible contamination, we can conduct detailed analyses to understand the origins of these flare events occurring in sdB/sdO binaries. We finally selected a refined sample consisting of 13 sdB/sdO stars, corresponding to 23 flare events (Table \ref{tab:flare_star_refined}).

Distributions comparing various flare properties and fractional flare luminosity between WDs and sdB/sdO stars are shown in Figure \ref{fig:stats}. Figure \ref{fig:stats}(a) displays the overall sample, while Figure \ref{fig:stats}(b) shows the subset with lowest pollution level from nearby objects (PL=0). We note that all compact stars in our sample show evidence of contamination from nearby objects or companion stars, substantially impacting the results. Despite pollution concerns, no obvious distinctions emerge in duration, amplitude or ED distributions when contrasting WD and sdB/sdO flares, although these distributions are more concentrated in Figure \ref{fig:stats}(b). Higher flare energies in sdB/sdO stars (Figure \ref{fig:stats}(a)) may link to contamination by nearby active stars, as energy calculations use stellar luminosity (see Eqn. \ref{eqn:etess}), which is greater in sdB/sdO stars. The diminishment of this discrepancy for the PL=0 subset in Figure \ref{fig:stats}(b) supports this contamination explanation. Furthermore, higher fractional flare luminosities for WDs could connect to their lower luminosity compared to sdB/sdO stars, allowing the flares from MS stars to more readily override and become detectable, and thus show higher level of flare activity. Again, this potential contamination effect diminishes when analyzing the PL=0 subset. Therefore, while definitive compact star flare attribution remains complex presently, comparing WD and sdB/sdO flare distributions provides initial insights on their distinct origins.

When we proceeded to analyze the FFDs, a notable observation emerged related to the power-law index $\alpha$ (see Figure \ref{fig:cffd}). We found that the FFDs from both sdB/sdO stars and WDs have an index $\alpha \sim 2$ when fitted with the entire flare sample, which is polluted by flares originated from other late-type MS stars (F-M type stars). These MS stars, with FFDs having an index $\alpha \sim 2$ \citep[e.g.,][]{Althukair_2023}, significantly skew the index $\alpha$ when fitted with the entire sample. However, when we move to the refined sample of sdB/sdO stars, we obtain an FFD that is less steep ($\alpha = 1.70$), which indicates a higher proportion of high energy flare events.

Before delving into potential explanations for this phenomenon, it is pertinent to note that several studies have investigated stellar flares across spectral types A to M. They found that the FFD from A-type stars has an index $1<\alpha<1.5$ when using Kepler photometry, a finding that markedly deviates from the $\alpha \sim 2$ observed in cooler F-M type stars \citep{Svanda_2016, Yang_2019, Bai_2020, Althukair_2023}. \citet{Yang_2023} also reported $\alpha = 1.76 \pm 0.19$ for A-type stars based on TESS data. Although this value is higher than that obtained via Kepler photometry, which is possibly due to increased contamination from lower angular resolution of TESS, it remains lower than indices for cooler MS stars. Such deviations in $\alpha$ have been interpreted as indicative of differing flare mechanisms between early-type (B/A-type) and late-type MS stars.

This observation prompts us to compare sdB/sdO stars with B/A-type MS stars, especially given the observed deviation in $\alpha$ between the FFDs derived from our entire sample and the refined sample of sdB/sdO stars. Notably, both sdB/sdO stars and B/A-type MS stars possess a radiative envelope, distinguishing them from cooler MS stars with convective envelopes. Despite the general expectation that these stars lack strong magnetic fields and are unlikely to produce flares through known dynamo mechanisms, as a result of the absence of a deep convective envelope \citep{Charbonneau_2010}, previous literature has reported that they might do show magnetic activities. For instance, \citet{Balona_2021} reported flare events on B/A-type MS stars and argued that observed rotational modulation in flaring B/A-stars suggests strong surface magnetic fields.

We thus hypothesize that some of the stellar flares from the refined sdB/sdO sample may originate solely on sdB/sdO itself, or through magnetic reconnection involving the sdB/sdO and its close companion. This hypothesis stems from the low incidence of superflares in late-type MS stars, the similar structure between hot MS stars and sdB/sdO stars, and most importantly, the decrease of the power-law index $\alpha$ from the FFD of the refined sdB/sdO sample compared to other samples. Analogous conclusions were previously proposed for B/A-type MS stars \citep{Balona_2021, Maryeva_2023}. We also propose stellar flares on sdB/sdO stars may operate through similar mechanisms to those detected in hot MS stars, if the flares in the refined sdB/sdO sample do arise exclusively from the sdB/sdO star. However, the nature of magnetic fields responsible for spots and flares in hot MS stars remains an open question. \citet{Svanda_2016} assumed these fields could be amplified by dynamo processes in the convective cores of A-type stars, subsequently becoming unstable and rising as magnetic ropes through the radiative envelope. \citet{Balona_2019} suggested differential rotation might suffice to generate local magnetic fields in B/A-type stars, presenting an exciting direction for future research to deepen our understanding of magnetic activities in stars with radiative envelopes. It is crucial to emphasize, however, that our hypothesis is preliminary and warrants further in-depth investigation.

We recall that, to our knowledge, there has been no dedicated survey searching for stellar flares in compact stars previously, although some searches have been conducted focusing on the outbursts in WDs \citep[see e.g.,][]{Bell_2016}. The lack of extensive surveys is largely due to challenges like complex light curve detrending \citep{Pietras_2022}. Our catalog could be the first step to such research, definitely, triggering new interests in stellar activity in highly evolved compact stars. The 13 sdB/sdO stars in the refined sample could be good candidates for future inspection. If confirmed, these candidates would be the first compact stars to exhibit a flare event. Moreover, while we cannot yet conclusively attribute any flares solely to an individual compact star, characterizing flare events in compact binary systems merits deeper investigation. For instance, \citet{Morgan_2016} demonstrated enhanced magnetic activity in M dwarfs with close WD companions compared to their isolated counterparts by analyzing their flare rates. Our results can help to examine such magnetic interactions across various compact binary systems.

We finally propose some prospects for our discoveries of flare events in sdB/sdO stars and WDs. Our methods are readily adaptable for similar analyses of Kepler and K2 photometry, which boast a higher angular resolution of roughly \qty{4}{arcsec.pixel^{-1}}, reducing contamination from nearby objects significantly. We anticipate the ongoing photometry from the second extension mission of TESS, which will further enable continuous monitoring of flare events in the entire sample. In addition, our method and pipeline can be helpful for flare hunting in other types of stars, or detecting other types of transient events in compact stars, with high confidence and feasibility. Furthermore, our tool, \texttt{tpfi}, is well-integrated with Kepler/K2 photometry, which enhances contamination reduction. Beyond the research on stellar flares, the identification charts generated by \texttt{tpfi} can also be invaluable for other studies, for instance, exoplanet detection and variable star research, underscoring its broader astronomical applicability.

\begin{acknowledgements}

We acknowledge the support from the National Natural Science Foundation of China (NSFC) through grants 12273002, 12090040, 12090042, 11988101 and 11933004. This work is supported by the International Centre of Supernovae at Yunnan Key Laboratory (Nos. 202302AN360001 and 202302AN36000102) and the science research grants from the China Manned Space Project. 
RS acknowledges support from INAF mini-grant on ``Hot subdwarfs and white dwarfs: Pulsations, Binaries and Planetary Systems". 
SC has financial support from the Centre National d'Etudes Spatiales (CNES, France). 
TC is supported by the LAMOST fellowship as a Youth Researcher, which is supported by the Special Funding for Advanced Users, budgeted and administrated by the Center for Astronomical Mega-Science, Chinese Academy of Sciences (CAMS), and acknowledges funding from the China Postdoctoral Science Foundation (2023M730297). 
TW thanks the supports from the National Key Research and Development Program of China (Grant No. 2021YFA1600402), the B-type Strategic Priority Program of Chinese Academy of Sciences (Grant No. XDB41000000), the Youth Innovation Promotion Association of Chinese Academy of Sciences and the Yunnan Ten Thousand Talents Plan Young \& Elite Talents Project.
All the TESS data used in this paper can be found in MAST at the TESS Light Curves - All Sectors HLSP \citep{10.17909/t9-nmc8-f686} and the TESS Target Pixel Files - All Sectors HLSP \citep{10.17909/t9-yk4w-zc73}.
The authors gratefully acknowledge the TESS team and all who have contributed to making this mission possible. Funding for the TESS mission is provided by the NASA Explorer Program.

\end{acknowledgements} 

\facility{TESS}
\software{astropy \citep{Astropy_2013, Astropy_2018, Astropy_2022},
          astroquery \citep{Astroquery},
          lightkurve \citep{Lightkurve},
          matplotlib \citep{Matplotlib},
          numpy \citep{Numpy},
          scipy \citep{Scipy}
}

\appendix

\section{Flare Events in the Refined Sample}

\begin{figure}
\epsscale{1.2}
\plotone{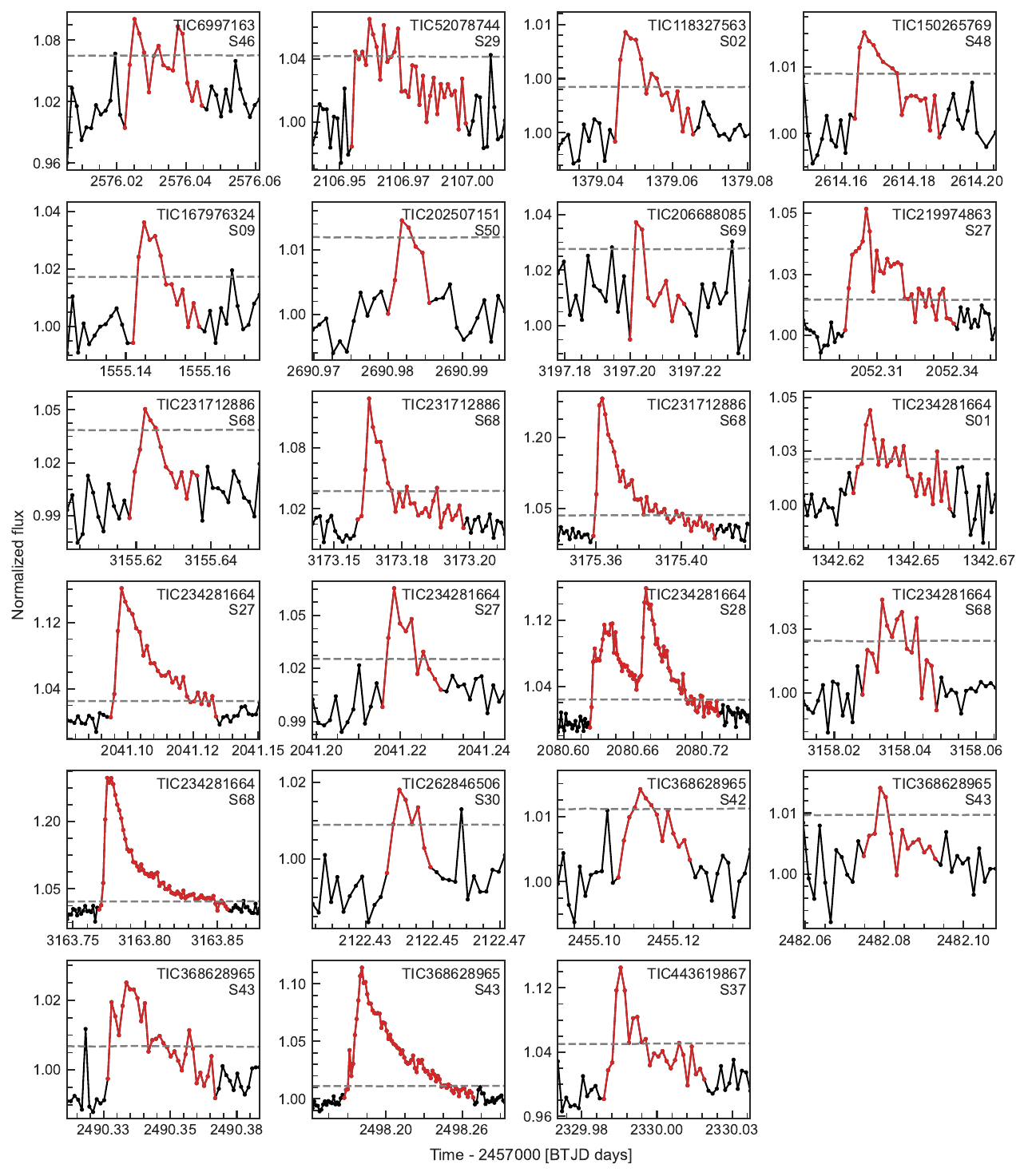}
\caption{All 23 flare events (red lines) in the light curves of the 13 sdB/sdO stars in our refined sample. The grey dashed lines are $3\sigma_{\rm MAD}$ from the median. TIC IDs and Sector numbers are indicated in the top right corner of each panel. The flare event in the light curve of TIC 234281664 in TESS Sector 28 is a complex flare event composed of two classical (single peak) flares.}
\label{fig:refined_flares}
\end{figure}

\bibliography{references}{}
\bibliographystyle{aasjournal}

\end{document}